# Accelerating CFD simulation with high order finite difference method on curvilinear coordinates for modern GPU clusters

Chuang-Chao Ye, Peng-Jun-Yi Zhang, Zhen-Hua Wan*, Rui Yan and De-Jun Sun

*Correspondence:
wanzh@ustc.edu.cn
Department of Modern Mechanics, University of Science and Technology of China, Hefei, China

**Abstract**

A high fidelity flow simulation for complex geometries for high Reynolds number (*Re*) flow is still very challenging, requiring a more powerful HPC system. However, the development of HPC with traditional CPU architecture suffers bottlenecks due to its high power consumption and technical difficulties. Heterogeneous architecture computation is raised to be a promising solution to the challenges of HPC development. GPU accelerating technology has been utilized in low order scheme CFD solvers on the structured grid and high order scheme solvers on unstructured meshes. The high-order finite difference methods on structured grids possess many advantages, e.g., high efficiency, robustness, and low storage. However, the strong dependence among points for a high-order finite difference scheme still limits its application on the GPU platform. In the present work, we propose a set of hardware-aware technology to optimize data transfer efficiency between CPU and GPU, as well as communication efficiency among GPUs. An in-house multi-block structured CFD solver with high order finite difference methods on curvilinear coordinates is ported onto the GPU platform and obtains satisfying performance with a speedup maximum of around 2000x over a single CPU core. This work provides an efficient solution to apply GPU computing in CFD simulation with specific high order finite difference methods on current GPU heterogeneous computers. The test shows that significant accelerating effects can be achieved for different GPUs.

**Keywords:** Hardware-aware, High order, Finite difference methods, Curvilinear coordinates, GPU

## 1 Introduction

Computational fluid dynamics (CFD) is one of the most important research methods in fluid mechanics, which highly relies on the computational capability of the computer, especially for accurately simulating realistic engineering flows. The deficiency of computational capability of computers has now become one of the biggest obstacles for future CFD development.

High order methods have been adopted in complex flow simulations like turbulence simulation, computational aeroacoustics (CAA) on the merit that they can obtain more accurate results with less grid than low order methods due to their low dissipation property. However, high order methods are more computationally expensive compared to low order methods. The real flows in nature are usually with large Reynolds numbers. Chapman estimated in 1979 the grid-point requirements ($N$) for wall-modeled large eddy simulation (LES) to be $N \sim Re_L^{2/5}$, for wall-resolving LES, $N \sim Re_L^{9/5}$, and for direct numerical simulation (DNS), $N \sim Re_L^{9/4}$ [1]. In 2012, Choi & Moin [2] revised Chapman's estimation to be $Re_L$, $Re_L^{13/7}$ and $Re_L^{37/14}$ respectively.

Since Gordon E. Moore, the co-founder of Intel, observed that the number of transistors on a microchip doubles every two years though the cost of computers is halved, Moore's law has governed the development of CPU for decades. However, in the past few years, there have been many voices about the gradual failure of this golden law because the high temperatures of transistors eventually would make it harder to create smaller circuits. In the meantime, there are many heterogeneous computational architectures arising, such as field-programmable gate array (FPGA), automata processor (AP), and graphics processing unit (GPU). Among those heterogeneous architectures, GPU computing has grown to be the most popular one in high-performance computing (HPC). GPU, originally designed for graphics rendering, releases great computational capability due to its "many cores" architecture. Compared to a traditional CPU which contains only several cores, GPU contains thousands of cores in a chip, which provides more powerful computation capability. In addition, GPU has lower power consumption and is cheaper for equivalent computation capability as CPU. In the latest TOP500 list of HPCs, half of the ten fastest HPC systems have been equipped with GPU devices. Summit, the fastest supercomputer in the world located in the Oak Ridge National Laboratory (ORNL) in the USA, consists of 4608 compute nodes, and each node contains 2 IBM POWER9 processors and 6 NVIDIA Volta V100 GPUs. The GPUs deliver over 95% performance of Summit. Compared to traditional CPU architecture supercomputer Tianhe-2A, Summit delivers double peak performance (200 PetaFlOPS) with nearly half power consumption (10KW).

The merits of GPU computing have attracted many attempts to port CFD simulation to the GPU platform. Tutkun, et al. [3] solve compressible Navier-Stokes equations with a compact scheme combining with a filter on a Tesla C1060 GPU and get a speedup of 9x to 16x compared to an AMD Phenom CPU. Esfahanian, et al. [4] perform 3rd and 5th order WENO scheme on two-dimensional uniform mesh solving Euler equations on GPU, and get a maximum speedup of 316x. Karantasis, et al. [5] perform 5th, 7th, and 9th order WENO scheme on a uniform grid with 6.9 million grid points, accelerating with Tesla S1070 GPU, and get a maximum speedup of 90x. Xu, et al. [6] perform high order finite difference WCNS scheme and Hybrid cell-edge and cell-node Dissipative Compact Scheme (HDCS) on curvilinear coordinates. The GPU-only approach achieves a speedup of about 1.3 on a Tesla M2050 GPU compared with two Xeon X5670 CPUs, while the hybrid CPU and GPU approach achieves a maximum 2.3x speedup. Lai, et al. [7, 8] develop a multiple GPU algorithm in hypersonic flow computations with the second-order finite volume method on curvilinear coordinates, and obtain 20x to 40x speedup accelerating with an Nvidia GTX1070 GPU compared to an eight cores Intel Xeon E2670 CPU. Elsen, et al. [9] develop a first-order to sixth-order finite difference method code for complex geometry on GPU platform, and obtain speedup 40x for a simple geometry and 20x for a complex geometry on a Nvidia 8800GTX GPU compared to an Intel-Core-2-Duo. Very recently, Lei, et al. [10] accelerated a compressible flow solver based on a second-order MUSCL/NND code on Cartesian coordinates with GPU. Tesla P100 GPU is used for accelerating and gets speedup 260x for MUSCL scheme and 144x for NND scheme over an E5-2640v4 CPU core. In general, for flow solvers in Cartesian coordinates and with low order schemes, the application of GPU gets conspicuous acceleration. But for a complicated coordinate system and high order schemes, the flow solvers get an inferior accelerating effect, which is far from the computational requirement for higher fidelity and larger-scale simulation.

High order methods based on unstructured mesh, e.g., flux reconstruction (FR) and discontinuous Galerkin (DG) methods, are popular in accurately simulating flows with complex geometry. GPU computing achieves great success in this branch of the high order method due to its computational independence of elements. Crabill, et al. [11] simulated the flow over a spinning golf ball at a Reynolds number of 150,000 with flux reconstruction (FR) solver on GPU. Witherden, et al. [12] developed a solver based on the high order flux reconstruction (FR), which has been used for a variety of flows. Although it is easy to improve precision, the numerical stability of this type of method in solving high-speed flows is poor, which limits its application. Furthermore, this method requires a large amount of storage, limiting its application in large-scale simulations in GPU computing. In contrast, the finite difference method is more storage saving and more robust when strong discontinuity exists. The finite difference method is still essential in high-order CFD simulations and deserves further development.

In the past few years, the hardware performance of GPU has been greatly improved, and a series of technologies have been developed, which make the heterogeneous computer more effective and mature. The current GPU computer is not simply a traditional CPU computer equipped with one or two GPUs, and it is a highly integrated hardware and software system. One computer has multiple GPUs, and GPUs become the main power of computing. To fully exploit the performance of current GPU computers, programs must adjust themselves to adapt to the complicated computer architectures.

In the present work, based on architecture analysis of modern GPU servers, we propose a set of hardware-aware technology to optimize the performance of data transfer between CPU and GPU , as well as the communication efficiency among GPUs. An in-house code is ported onto GPU with careful memory planning and high-efficiency kernel design. Moreover, a modified alternative formulation of the WENO scheme, which saves computation and is more efficient in GPU computing, is proposed. The rest of the paper is organized as follows. The numerical methods are presented in Section 2. Then, the hardware facility is introduced in Section 3. The programming and optimization strategies are introduced in detail in Section 4. Furthermore, the performance cases and practical case are presented in Section 5 and 6, respectively. Finally, some conclusions are summarized in Section 7.

## 2 Numerical methods

The flow is governed by nondimensional Navier-Stokes equations

$$
\begin{aligned}
&\frac{\partial \rho}{\partial t}+\frac{\partial \rho u_j}{\partial x_j}=0\\
&\frac{\partial \rho u_j}{\partial t}+\frac{\partial \rho u_i u_j}{\partial x_j}=-\frac{\partial p}{\partial x_i}+\frac{\partial \rho \tau_{ij}}{\partial x_j}\\
&\frac{\partial e}{\partial t}+\frac{\partial (e+p)u_j}{\partial x_j}=-\frac{\partial q_j}{\partial x_j}+\frac{\partial u_i \tau_{ij}}{\partial x_j}
\end{aligned}
\tag{1}
$$

where

$$T=\frac{\gamma p}{\rho},$$

is the temperature,

$$e=\frac{p}{\gamma-1}+\frac{1}{2}\rho(u^2+v^2+w^2)$$

is the total energy per unit volume,

$$\tau_{ij}=\frac{Ma_\infty}{Re_\infty}\left[\mu\left(\frac{\partial u_i}{\partial x_j}+\frac{\partial u_j}{\partial x_i}-\frac{2}{3}\frac{\partial u_k}{\partial x_k}\delta_{ij}\right)\right]$$

is the viscous stress tensor, and

$$q_j=-\left[\frac{Ma_\infty \mu}{Re_\infty Pr_\infty(\gamma-1)}\right]\frac{\partial T}{\partial x_i}$$

is the heat flux vector. In the above equations, $\rho$ denotes density, $u_j$ are velocity components, $p$ is the pressure, and $T$ denotes temperature. $\gamma = 1.4$ is the specific heat ratio. $Ma_\infty$ and $Re_\infty$ are free-stream Mach number and Reynolds number, respectively. The present study is based on an inhouse code "***HiResX*** (**Hi**gh **Res**olution flow **S**olver)", which aims to simulate compressible flows with complex geometries in high order accuracy with finite difference methods. To treat complex geometries, we solve the Navier-Stokes equations on curvilinear coordinates, which can be written as

$$\frac{\partial \tilde{U}}{\partial t}+\frac{\partial(\tilde{E}-\tilde{E}_v)}{\partial \xi}+\frac{\partial(\tilde{F}-\tilde{F}_v)}{\partial \eta}+\frac{\partial(\tilde{G}-\tilde{G}_v)}{\partial \zeta}=0 \tag{2}$$

where $\widetilde{U}$ is the vector of conservative flow variables, $\widetilde{E}, \widetilde{F}, \widetilde{G}$ are inviscid flux vectors, and $\widetilde{E_v}, \widetilde{F_v}, \widetilde{G_v}$ are viscous flux vectors, defined as

$$\widetilde{U}=\frac{1}{J}\Big(\rho, \rho u, \rho v, \rho w, e\Big)^T \tag{3}$$

$$\widetilde{E}=\frac{1}{J}\begin{pmatrix}\rho U\\ \rho u U+\xi_x p\\ \rho v U+\xi_y p\\ \rho w U+\xi_z p\\ (e+p)U\end{pmatrix}, \widetilde{F}=\frac{1}{J}\begin{pmatrix}\rho V\\ \rho u V+\eta_x p\\ \rho v V+\eta_y p\\ \rho w V+\eta_z p\\ (e+p)V\end{pmatrix}, \widetilde{G}=\frac{1}{J}\begin{pmatrix}\rho W\\ \rho u W+\zeta_x p\\ \rho v W+\zeta_y p\\ \rho w W+\zeta_z p\\ (e+p)W\end{pmatrix}, \tag{4}$$

$$\widetilde{E_v}=\frac{1}{J}\begin{pmatrix}0\\ \xi_x\tau_{xx}+\xi_y\tau_{xy}+\xi_z\tau_{xz}\\ \xi_x\tau_{yx}+\xi_y\tau_{yy}+\xi_z\tau_{yz}\\ \xi_x\tau_{zx}+\xi_y\tau_{zy}+\xi_z\tau_{zz}\\ \xi_x\beta_x+\xi_y\beta_y+\xi_z\beta_z\end{pmatrix}, \widetilde{F_v}=\frac{1}{J}\begin{pmatrix}0\\ \eta_x\tau_{xx}+\eta_y\tau_{xy}+\eta_z\tau_{xz}\\ \eta_x\tau_{yx}+\eta_y\tau_{yy}+\eta_z\tau_{yz}\\ \eta_x\tau_{zx}+\eta_y\tau_{zy}+\eta_z\tau_{zz}\\ \eta_x\beta_x+\eta_y\beta_y+\eta_z\beta_z\end{pmatrix}, \widetilde{G_v}=\frac{1}{J}\begin{pmatrix}0\\ \zeta_x\tau_{xx}+\zeta_y\tau_{xy}+\zeta_z\tau_{xz}\\ \zeta_x\tau_{yx}+\zeta_y\tau_{yy}+\zeta_z\tau_{yz}\\ \zeta_x\tau_{zx}+\zeta_y\tau_{zy}+\zeta_z\tau_{zz}\\ \zeta_x\beta_x+\zeta_y\beta_y+\zeta_z\beta_z\end{pmatrix} \tag{5}$$

$$\begin{gathered}\tau_{xx}=\lambda(u_x+v_y+w_z)+2\mu u_x,\tau_{xy}=\tau_{yx}=\mu(u_y+v_x)\\ \tau_{yy}=\lambda(u_x+v_y+w_z)+2\mu v_y,\tau_{xz}=\tau_{zx}=\mu(u_z+w_x)\\ \tau_{zz}=\lambda(u_x+v_y+w_z)+2\mu w_z,\tau_{yz}=\tau_{zy}=\mu(v_z+w_y)\\ \beta_x=u\tau_{xx}+v\tau_{xy}+w\tau_{xz}+\dot{q}_x\\ \beta_y=u\tau_{xy}+v\tau_{yy}+w\tau_{yz}+\dot{q}_y\\ \beta_z=u\tau_{xz}+v\tau_{yz}+w\tau_{zz}+\dot{q}_z\\ \dot{q}_x=\frac{\mu}{Pr\,(\gamma-1)}\frac{\partial T}{\partial x},\dot{q}_y=\frac{\mu}{Pr\,(\gamma-1)}\frac{\partial T}{\partial y},\dot{q}_z=\frac{\mu}{Pr\,(\gamma-1)}\frac{\partial T}{\partial z}\end{gathered} \tag{6}$$

where $(\xi,\eta,\zeta)_{x,y,z}$ are metrics, and $J$ is the coordinate transformation Jacobian. To minimize the discretization error, we compute the metrics and Jacobian in a conservative form, defined as:

$$\begin{cases}\widehat{\xi}_x=\xi_x/J=(y_\eta z)_\zeta-(y_\zeta z)_\eta,\widehat{\xi}_y=\xi_y/J=(z_\eta x)_\zeta-(z_\zeta x)_\eta,\widehat{\xi}_z=\xi_z/J=(x_\eta y)_\zeta-(x_\zeta y)_\eta\\ \widehat{\eta}_x=\eta_x/J=(y_\zeta z)_\xi-(y_\xi z)_\zeta,\widehat{\eta}_y=\eta_y/J=(z_\zeta x)_\xi-(z_\xi x)_\zeta,\widehat{\eta}_z=\eta_z/J=(x_\zeta y)_\xi-(x_\xi y)_\zeta\\ \widehat{\zeta}_x=\zeta_x/J=(y_\xi z)_\eta-(y_\eta z)_\xi,\widehat{\zeta}_y=\zeta_y/J=(z_\xi x)_\eta-(z_\eta x)_\xi,\widehat{\zeta}_z=\zeta_z/J=(x_\xi y)_\eta-(x_\eta y)_\xi\end{cases} \tag{7}$$

$$J^{-1}=x_\xi y_\eta z_\zeta-x_\xi y_\zeta z_\eta+x_\eta y_\zeta z_\xi-x_\eta y_\xi z_\zeta+x_\zeta y_\xi z_\eta-x_\zeta y_\eta z_\xi \tag{8}$$

Here, the derivatives in metrics and Jacobian, and also the derivatives of viscous terms are discretized by the sixth order central scheme.

The derivatives of inviscid fluxes are approximate with the high order alternative formulation of weighted essentially non-oscillatory scheme (AFWENO) [13]. Different from the reconstruction idea of the traditional high-order WENO scheme [14], this new formulation of WENO is based on the idea of interpolation. Similar to the traditional WENO scheme, the derivative is written in a conservative form:

$$\frac{df(u)}{dx}=\frac{1}{\Delta x}\left(\hat{f}_{i+\frac{1}{2}}-\hat{f}_{i-\frac{1}{2}}\right) \tag{9}$$

The numerical fluxes are approximated as

$$\hat{f}_{i+\frac{1}{2}}=f_{i+\frac{1}{2}}+\sum_{k=1}^{[(r-1)/2]}a_{2k}\Delta x^{2k}\left.\frac{\partial^{2k}f}{\partial x^{2k}}\right|_{i+\frac{1}{2}}+O\left(\Delta x^{r+1}\right), \tag{10}$$

where the coefficients $a_{2k}$ can be obtained by Taylor expansion. In order to achieve fifth order accuracy, we here fix $r=5$ and then

$$\hat{f}_{i+\frac{1}{2}}=f_{i+\frac{1}{2}}-\frac{1}{24}\Delta x^2 f_{xx}\Big|_{i+\frac{1}{2}}+\frac{7}{5760}\Delta x^4 f_{xxxx}\Big|_{i+\frac{1}{2}}, \tag{11}$$

where $f_{i+\frac{1}{2}}$ is the numerical flux at $x_{i+\frac{1}{2}}$, which can be approximated as

$$f_{i+\frac{1}{2}}=h\left(u^-_{i+\frac{1}{2}},u^+_{i+\frac{1}{2}}\right). \tag{12}$$

The values of $u^{\pm}_{i+\frac{1}{2}}$ can be obtained with a linear combination of the low order interpolation approximate values based on the small stencils, with the global stencil biases to left for " + ", and right bias for " − ".

$$u_{i+\frac{1}{2}}=w_1u^{(1)}_{i+\frac{1}{2}}+w_2u^{(2)}_{i+\frac{1}{2}}+w_3u^{(3)}_{i+\frac{1}{2}} \tag{13}$$

The approximations based on three sub-stencils are given explicitly as

$$\begin{aligned} u^{(1)}_{i+\frac{1}{2}} &= \frac{3}{8}u_i + \frac{3}{4}u_{i+1} - \frac{1}{8}u_{i+2} \\ u^{(2)}_{i+\frac{1}{2}} &= -\frac{1}{8}u_{i-1} + \frac{3}{4}u_i + \frac{3}{8}u_{i+1} \\ u^{(3)}_{i+\frac{1}{2}} &= \frac{3}{8}u_{i-2} - \frac{5}{4}u_{i-1} + \frac{15}{8}u_i \end{aligned} \tag{14}$$

The form of nonlinear weights of WENO reads

$$w_j = \frac{\tilde{w}_j}{\tilde{w}_1 + \tilde{w}_2 + \tilde{w}_3}, \quad \text{with} \quad \tilde{w}_j = \frac{\gamma_j}{\left(\epsilon + \beta_j\right)^2} \tag{15}$$

where $\gamma_j$ $(j = 1, 2, 3)$ are the ideal linear weights, given as

$$\gamma_1 = \frac{1}{16}, \quad \gamma_2 = \frac{5}{8}, \quad \gamma_3 = \frac{5}{16} \tag{16}$$

$\epsilon$ is a small positive number to avoid the denominator becoming zero, which is typically fixed to be $\epsilon = 10^{-6}$ in actual computations. The smoothness indicators are chosen as the scaled sum of the square $L^2$ norms of all the derivatives of interpolation polynomial, which can be easily worked out as

$$\begin{aligned} \beta_1 &= \frac{1}{3}\left(4u^2_{i-2} - 19u_{i-2}u_{i-1} + 25u^2_{i-1} + 11u_{i-2}u_i - 31u_{i-1}u_i + 10u^2_i\right) \\ \beta_2 &= \frac{1}{3}\left(4u^2_{i-1} - 13u_{i-1}u_i + 13u^2_i + 5u_{i-1}u_{i+1} - 13u_iu_{i+1} + 4u^2_{i+1}\right) \\ \beta_3 &= \frac{1}{3}\left(10u^2_i - 31u_iu_{i+1} + 25u^2_{i+1} + 11u_iu_{i+2} - 19u_{i+1}u_{i+2} + 4u^2_{i+2}\right) \end{aligned} \tag{17}$$

The two-argument function $h$ is a monotone flux, which can be chosen from exact or any approximate Riemann solvers, such as HLL, and Roe. In this work, the Steger-Warming flux is utilized to construct the numerical flux $f_{i+\frac{1}{2}}$.

The second and fourth order derivatives in Eq. (11) can be simply explicitly approximated with a central scheme as

$$\begin{aligned} f_{xx}\big|_{i+\frac{1}{2}} &= \frac{1}{48}(-5f_{i-2} + 39f_{i-1} - 34f_i - 34f_{i+1} + 39f_{i+2} - 5f_{i+3}) \\ f_{xxxx}\big|_{i+\frac{1}{2}} &= \frac{1}{2}(f_{i-2} - 3f_{i-1} + 2f_i + 2f_{i+1} - 3f_{i+2} + f_{i+3}) \end{aligned} \tag{18}$$

Substitute Eqs.(11) and (18) into Eq. (9), and we get the final expression as

$$f_x\big|_i = f_{i+\frac{1}{2}} - f_{i-\frac{1}{2}} + \frac{17}{256}(f_{i+1} - f_{i-1}) - \frac{13}{320}(f_{i+2} - f_{i-2}) + \frac{19}{3840}(f_{i+3} - f_{i-3}) \tag{19}$$

The semi-discretized Eq. (9) can be explicitly solved with the 3-stages TVD Runge-Kutta scheme as follows

$$\begin{aligned} Q^{(1)} &= Q^n + \Delta t\, R\left(Q^n\right) \\ Q^{(2)} &= \frac{3}{4}Q^n + \frac{1}{4}Q^{(1)} + \frac{1}{4}\Delta t\, R\left(Q^{(1)}\right) \\ Q^{n+1} &= \frac{1}{3}Q^n + \frac{2}{3}Q^{(2)} + \frac{2}{3}\Delta t\, R\left(Q^{(2)}\right) \end{aligned} \tag{20}$$

## 3 Hardware environment

A traditional CPU-based supercomputing cluster usually consists of hundreds of thousands of computational nodes, and each node contains dozens of cores. These nodes are interconnected with high speed networks. Up to now, the number of cores in a single CPU is still limited. Due to this limitation, the large task must be divided into a lot of partitions and distributed to many nodes. One obvious drawback is that the complex communication between nodes may lead to high latency. Furthermore, the price/performance ratio of traditional CPU-based cluster is low, due to the fact that you have to pay for the frame of each computation node by increasing CPU cores, which does not directly contribute to computational capabilities.

The computational capability of a computing device with specific space occupation can be represented as "computing density". In the past few years, supercomputing has come to the heterogeneous computing era, and higher computing density supercomputers have been developed. Those computers are accelerated with some devices called many-core processor accelerators, such as GPU and FPGA. In the realm of heterogeneous computing, GPU is the most popular one. Figure 1 shows the framework of the modern GPU server cluster. Each node contains two CPUs and much more GPUs than before. The GPUs in the same server communicate with each other through PCIe slots and further communicate with GPUs in other servers by the network. A more sophisticated technology called Nvlink has been developed by NVIDIA to enable direct data transfer between GPUs in the same server at a faster speed than PCIe's.

The two GPU servers utilized in this work are equipped with 10 NVIDIA RTX 2080 Ti GPUs and 10 NVIDIA Titan V GPUs, respectively. The RTX 2080 Ti GPU utilizes the newest Turing architecture of NVIDIA GPU, while Titan V utilizes the architecture Volta. The RTX 2080 Ti GPU is a consumer-oriented product aiming at video and games, though it has up to 4352 CUDA cores, the double-precision (DP) performance is only 1/32 of single-precision (SP) performance. However, the core GV100 of Titan V contains 5120 CUDA cores, and the ratio of DP/SP is up to 1/2, which makes it still the most powerful GPU core up to now. We also utilize the Tesla P100 GPU, based on much older architecture Pascal. Although it contains fewer CUDA cores than RTX 2080 Ti, it has a much higher theoretical DP performance. The detailed specs of GPUs mentioned above can be found in Table 1. The CPU of the server is Intel Xeon E5 2680 v3, with 12 cores

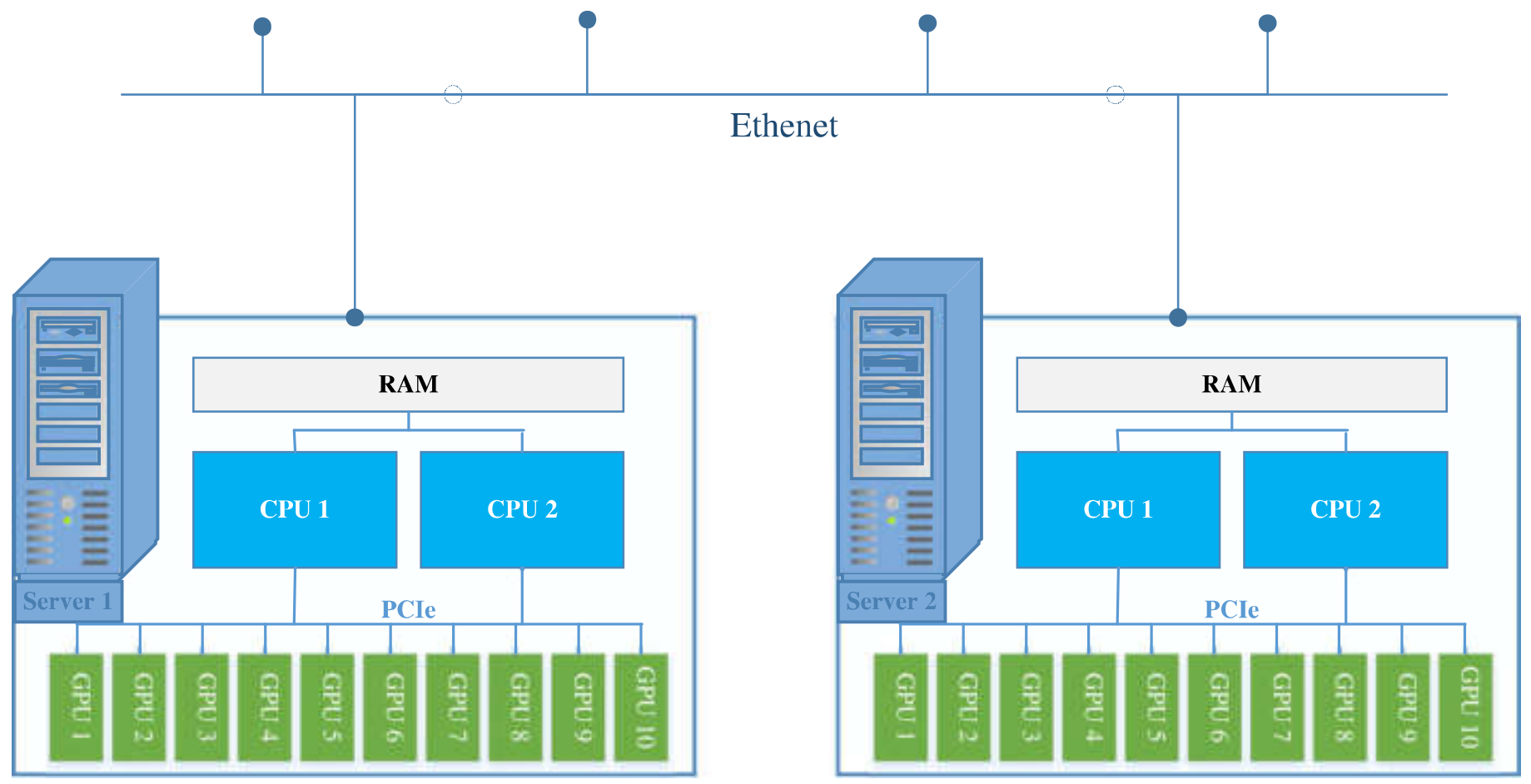

**Fig. 1** Framework of modern GPU cluster. The computational nodes are connected with high speed networks. GPUs deliver the majority of performance of modern GPU cluster

**Table 1** Present GPU specs

| GPU Models | RTX 2080 Ti | Tesla P100 | Titan V |
|---|---|---|---|
| CUDA Cores | 4352 | 3584 | 5120 |
| SM count | 68 | 56 | 80 |
| Based Clock (MHz) | 1350 | 1190 | 1200 |
| Total Memory | 11 GB GDDR6 | 16 GB HBM2 | 12 GB HBM2 |
| Memory Bandwidth (GB/sec) | 616 | 732.2 | 652.8 |
| FP32 Performance (TFlops) | 13.45 | 9.3 | 14.90 |
| FP64 Performance (TFlops) | 0.42 | 4.7 | 7.45 |
| Computation Capability | 7.5 | 6.0 | 7.0 |
| Architecture | Turing | Pascal | Volta |

Titan V has the highest double precision operation performance. RTX 2080 Ti utilizes the newest architecture and provides the highest single precision operation performance in a single CPU, and the spec is given in Table 2. The ideal computational performance of dual-CPU in the server is 960 GFlops (FP64 performance). With GPU acceleration, 10 RTX 2080 Ti GPUs deliver 134.5 TFlops SP float performance and 4.2 TFlops DP float performance, while 10 Titan V GPUs deliver an incredible 149.0 TFlops SP performance and 74.5 TFlops DP performance. In this work, no Nvlink is applied.

## 4 Programming implementation and optimization of *HiResX*

### 4.1 Code introduction

***HiResX*** is originally written in Fortran 90, aiming at simulating various compressible flows with complex domains in high order schemes. ***HiResX*** solver is equipped with various turbulence models such as Reynolds-Averaged Navier-Stokes (RANS) models, LES model, and detached eddy simulation (DES) model to simulate realistic engineering flows with high Reynolds numbers. Message Passing Interface (MPI) is utilized to distribute computation tasks to different CPU cores, inter- or intra- nodes.

There are several strategies for performing computation on GPU devices. OpenACC offers a user-driven lightweight solution that the user needs to simply add some directive-based clauses in the code segment, which needs to accelerate on GPU with significantly less programming effort than required with a low-level model. CUDA Fortran based on PGI Fortran is another solution provided by PGI company, which offers programming flexibility using Fortran language. However, CUDA Fortran highly relies on the PGI compiler. CUDA C/C++ is a programming language based on standard C/C++ and developed by NVIDIA, and it provides better affinity and flexibility to operate NVIDIA's GPUs at low-level programming. However, programming with CUDA C/C++ needs much higher

**Table 2** Present spec of CPU

| Intel Xeon E5 2680 v3 | |
|---|---|
| # of Cores | 12 |
| # of thread | 24 |
| Based Frequency (GHz) | 2.50 |
| Cache | 30 MB |
| Max Memory Bandwidth | 68 GB/sec |
| FP32 Performance | 960 GFlops |
| FP64 Performance | 480 GFlops |

The double precision operation performance of E5-2680v3 is slightly higher than RTX 2080 Ti's skill. To port an existing complex code onto GPU by completely rewriting it with another programming language is not a wise choice. In order to fully exploit the performance of the current GPU server without rewriting the whole solver, mixed programming of Fortran and C/C++ is applied, and GPU computing is achieved with CUDA C/C++.

GPU computing for each process is independent with each other when ***HiResX*** is running in parallel, thus ***HiResX*** supports three parallel modes: CPU-only mode, GPU-only mode and collaborating (or hybrid) CPU/GPU mode. In CPU-only parallel mode, all processes perform computation on CPU only, whether the computation node has GPU devices or not. This parallel mode is suitable for most of the traditional supercomputers, which support CPU computing only. In GPU-only parallel mode, all processes are running on GPU devices. Hence for a supercomputing cluster, every computational node running the code must be equipped with a GPU device. The hybrid CPU/GPU mode is a parallel mode that can make full use of all CPU and GPU resources in the cluster, as mentioned in Ref. [15]. For this parallel mode, the computational capability of CPU and GPU in one node should be roughly comparative.

The pseudo code and procedures of GPU computing of ***HiResX*** are illustrated in Figs. 2 and 3. After the initialization of the program, the process which runs computation on GPU uploads the conservative variables $Q$, primitive variables $Q_v$, Jacobian $J$, and metrics at cell node to GPU device. To avoid frequent data transfer between CPU and GPU, all these variables in the computation domain are uploaded to GPU and stored in GPU's global memory. The Jacobian $J$ and metrics at the cell edge, which are computed by CPU at startup, would not be uploaded with consideration of three aspects. Firstly, if all these variables at cell edges in three directions are stored in the memory of GPU, additional

```
initialize(...)  // Initialize solver
loop iter → maximum iterations
    if iGPU is Enabled
        timestep_gpu(...)  // Loop of simulation
    else
        timestep_cpu(...)  // Loop of simulation
    end if
    loop i → number of RK stages
        if iGPU is Enabled
            cflux_gpu(...)  // Compute inviscid flux
            vflux_gpu(...)  // Compute viscid flux
            time_advance_gpu(...)  // time advance with R-K
            bc_block_gpu(...) // Exchange information near block boundary
            update_vars_gpu(...) //Update primitive variables
            bc_phys_gpu(...) //Update variables at physical boundary
        else
            cflux_cpu(...)  // Compute inviscid flux
            vflux_cpu(...)  // Compute viscid flux
            time_advance_cpu(...)  // time advance with R-K
            bc_block_cpu(...) // Exchange information near block boundary
            update_vars_cpu(...) //Update primitive variables
            bc_phys_cpu(...) //Update variables at physical boundary
        end if
    end loop number of RK stages
    if iGPU is Enabled
        calc_res_gpu(...)  // Compute reisidual
    else
        calc_res_cpu(...)  // Compute reisidual
    end if
end loop maximum iterations
finalize(...) // End of program
```

**Fig. 2** Main structure of ***HiResX*** summarized by pseudo code

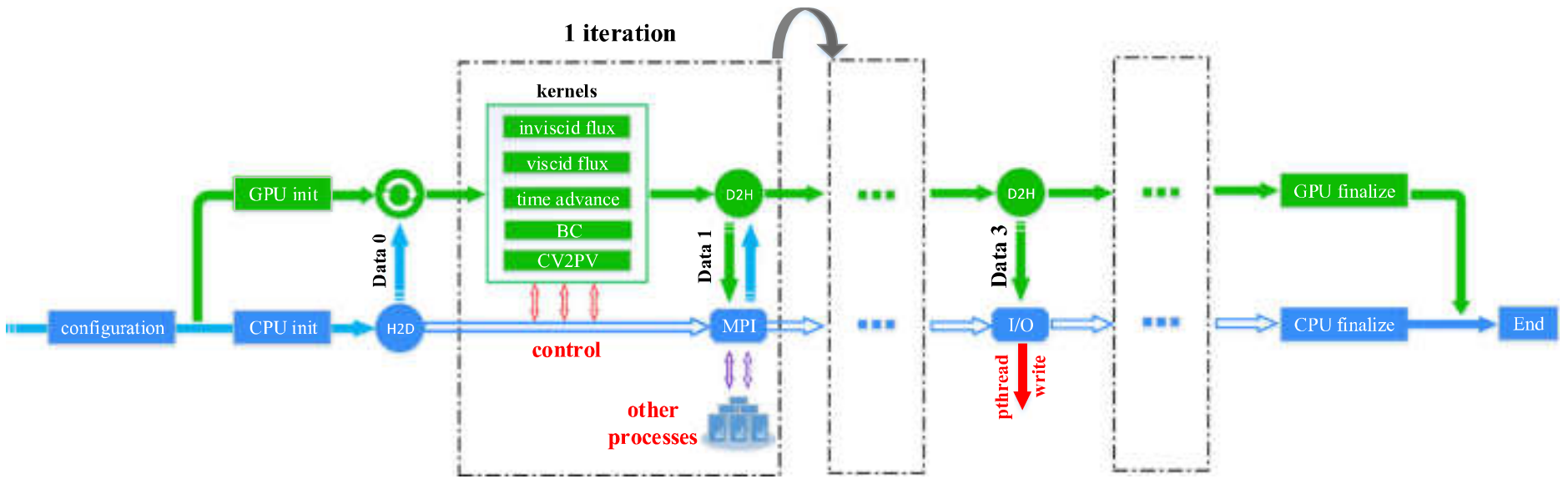

**Fig. 3** A global glance of execution procedures of ***HiResX*** running on GPU. Except for initialization and data performed on CPU, all computations are performed on GPU while CPU is used to schedule kernels of GPU only. "Data 0" represents the initial data computed in CPU and uploaded to GPU. "Data 1" represents data exchanged between processes. "Data 3" represents flowfield data to be written, and it should be downloaded to CPU and written with a new thread

memory space for 12 variables is needed, which reduces the maximum grid number that a single GPU can deal with. Secondly, if only memory space of one direction is allocated on GPU, though the memory space is reduced to a third of the former case, frequent data transfer during direction switching is time-consuming. Thirdly, for GPU, computational capability is much stronger than memory bandwidth. Hence direct interpolation from Jacobian and metrics at cell node is more time-saving. The calculation of residual $dQ$ on the right-hand side, time advancement, and primitive variables updating from conservative variables is completely running on GPU. When physical boundary condition is applied, very little data needs to be uploaded onto GPU, such as the ID number of boundary points, which takes very little time that can be ignored.

For traditional CPU computers, the user may not concern about where and how their program is running. When the program runs in parallel, the user needs only to distribute the computation to different CPU processes with MPI or to different threads by tools (e.g., OpenMP). Users rarely care about the hardware information when developing the program. However, in current GPU computing, data exchanges at connecting block boundaries are more complex due to the complex environment in heterogeneous computational architecture. Figure 4 shows the communication structures of our solver. When the solver runs in CPU-only mode, data exchange among blocks in the same process is straightforward by simply copying data without communication with other processes, as is depicted in the figure with the yellow pentagon pairs. For blocks in different processes in the same node, data exchange among them is achieved by the use of shared memory communication technique in MPI-3 standard, as is depicted in the figure with the red five-pointed star pairs. For blocks in different processes in different nodes, data exchange is achieved by standard MPI communication, which is marked as the brown triangle pairs.

In GPU computing, the data must be uploaded to GPU from CPU first and downloaded back to CPU when GPU computation is finished. The data has to be transferred between memories of CPU and GPU through the PCIe bus. In the early stage of the development of GPU computing, each GPU is installed in different PCIe slots and works independently, which communicates with the CPU only. If the program is running in parallel with multiple processes on GPUs, the data to be sent has to be downloaded from GPU, and then communication is fully performed in CPU through MPI when communications among processes are involved. After the data is exchanged, the received data has to be uploaded

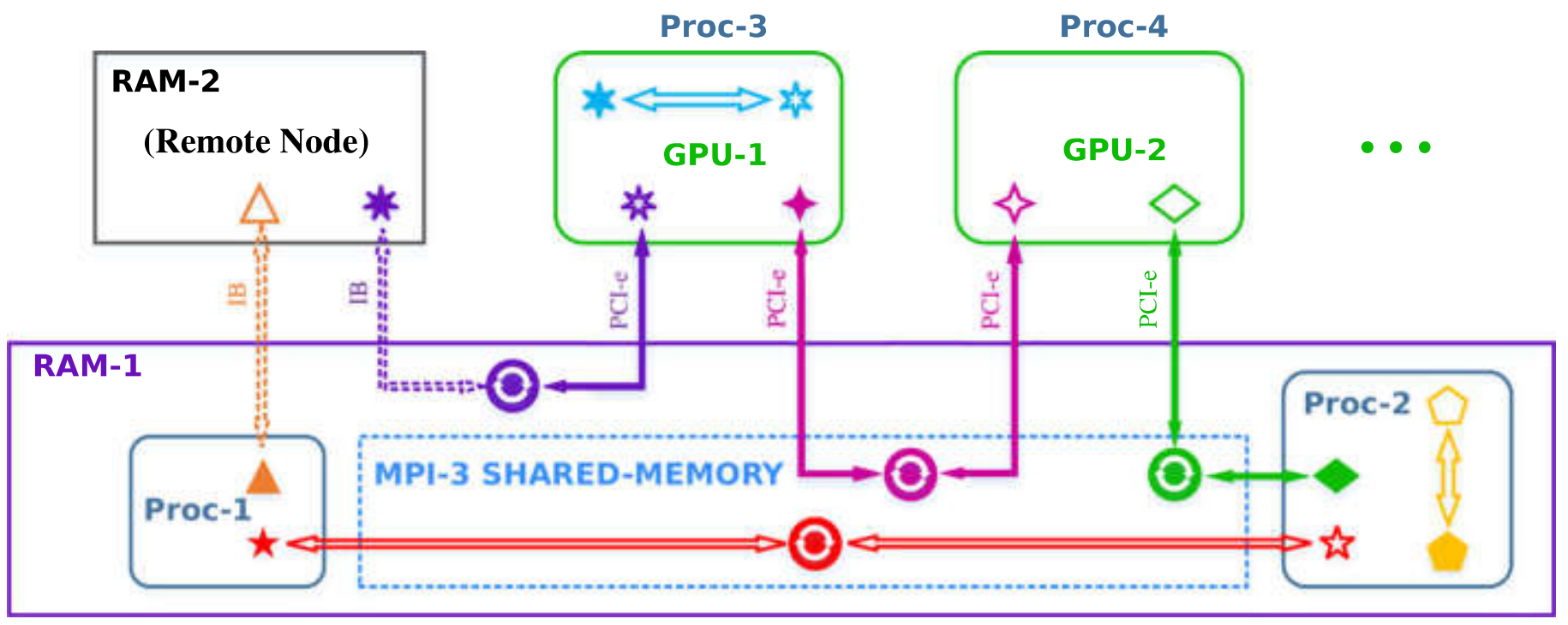

**Fig. 4** Framework of communication of ***HiResX*** without GPU peer to peer communication technology support. For communication between processes that are all running on CPU, if processes are located in the same node, they exchange data within RAM with MPI-3 shared memory technique, see red path. If processes are located in different nodes, the standard MPI communication is utilized, see yellow path that connects triangle pair. For processes with GPU acceleration, data on GPU should be downloaded back to CPU, and then processes communicate in the same way as processes without acceleration

to GPU. It is obvious in this procedure that processes in different GPUs communicate by use of CPU memory and explicit data copying between CPU and GPU. When data on GPU is downloaded to CPU, the communication restores to general communication in CPU, and all techniques can be applied, as is depicted with four-pointed star pair and seven-pointed star pair. However, the Unified Virtual Addressing (UVA) technology and CUDA-aware MPI technology make this procedure simpler and more efficient. There is no need to perform explicit data copying because the MPI interfaces can recognize the location of buffer data and then find the optimal path to transfer data. For blocks in the same process with GPU accelerating, data can be exchanged on GPU without leaving the device.

When the program runs in hybrid CPU/GPU mode, the communications among processes without GPU accelerating (CPU-only mode) are the same as the communications among processes with GPU accelerating (GPU-only mode). The only thing to deal with is the communication between processes with or without GPU accelerating, as is marked with diamond pair in Fig. 4. For the process with accelerating, data to be sent should be downloaded to CPU memory explicitly if CUDA-aware MPI is not applied, while the process without accelerating does nothing with GPU. When data is downloaded to the CPU, the communication can be performed similarly to CPU-only mode. However, with CUDA-aware MPI technology and UVA, the data in GPU and CPU can be exchanged directly.

### 4.2 Domain decomposition

The ***HiResX*** solver is developed to simulate flows with complex domains with a multi-block structure grid. Figure 5 illustrates the domain decomposition strategies of ***HiResX***. The whole computation domain is split into many connecting blocks, and these blocks are distributed to different processors. For hybrid CPU/GPU parallel computing, the sizes of grid blocks can be different. The grid block size for CPU is usually much smaller than that for GPU since the latter has higher computational capability.

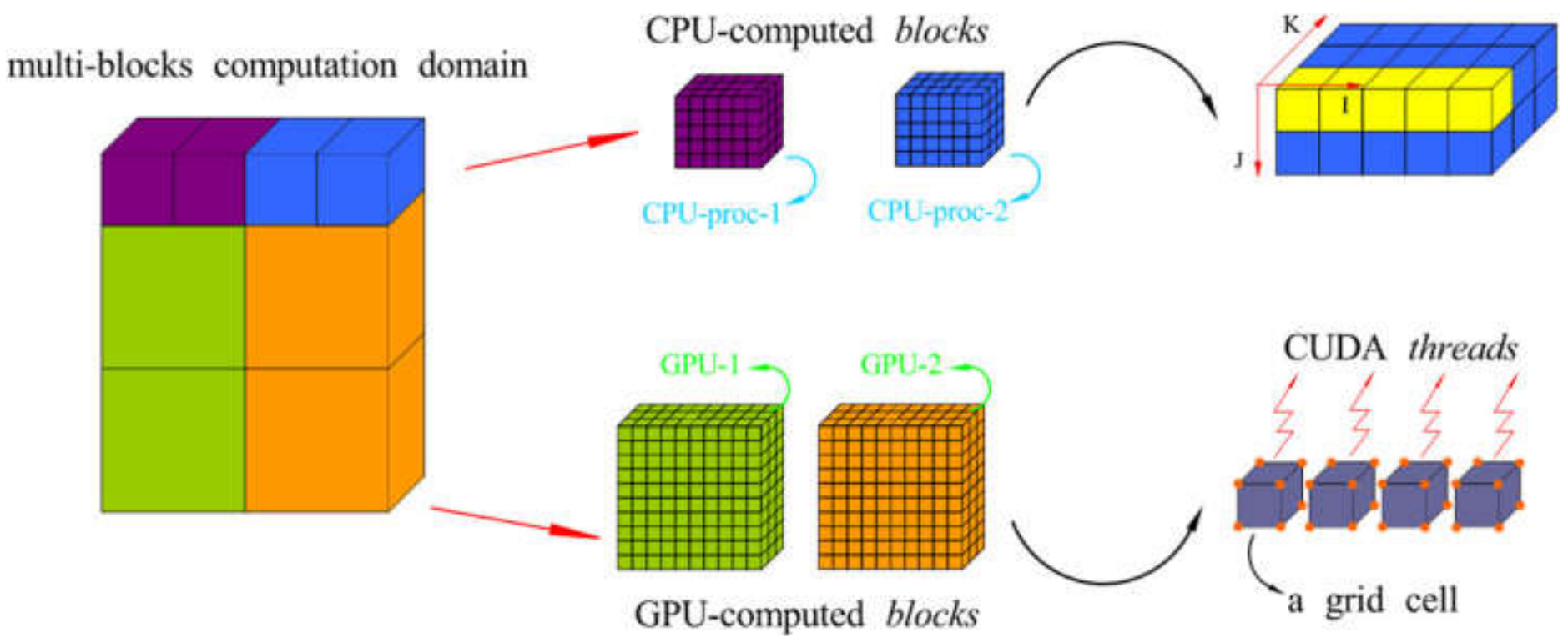

**Fig. 5** Domain decomposition strategies of ***HiResX***. The domain is divided into several blocks according to performance of CPU and GPU in order to balance workloads of each process. For the blocks computed by CPU, the whole block is computed by a CPU process. For the blocks computed by GPU, each grid cell is computed by one CUDA thread

For ***HiResX***, in order to balance the computational loads of processes of CPUs and GPUs, the whole domain can be split arbitrarily without the limitation of one-to-one block connection, and the block connecting faces and physical boundary faces can also be arbitrarily defined on blocks.

### 4.3 Hardware-aware technique

In the current GPU server, there are more GPUs installed in a single machine. Moreover, a variety of PCIe root architectures have been designed for different performance needs, as demonstrated in Fig. 6. The whole hardware system needs to match the increasing performance of current GPUs, and the solver also needs to adapt to the hardware system for maximum computational performance.

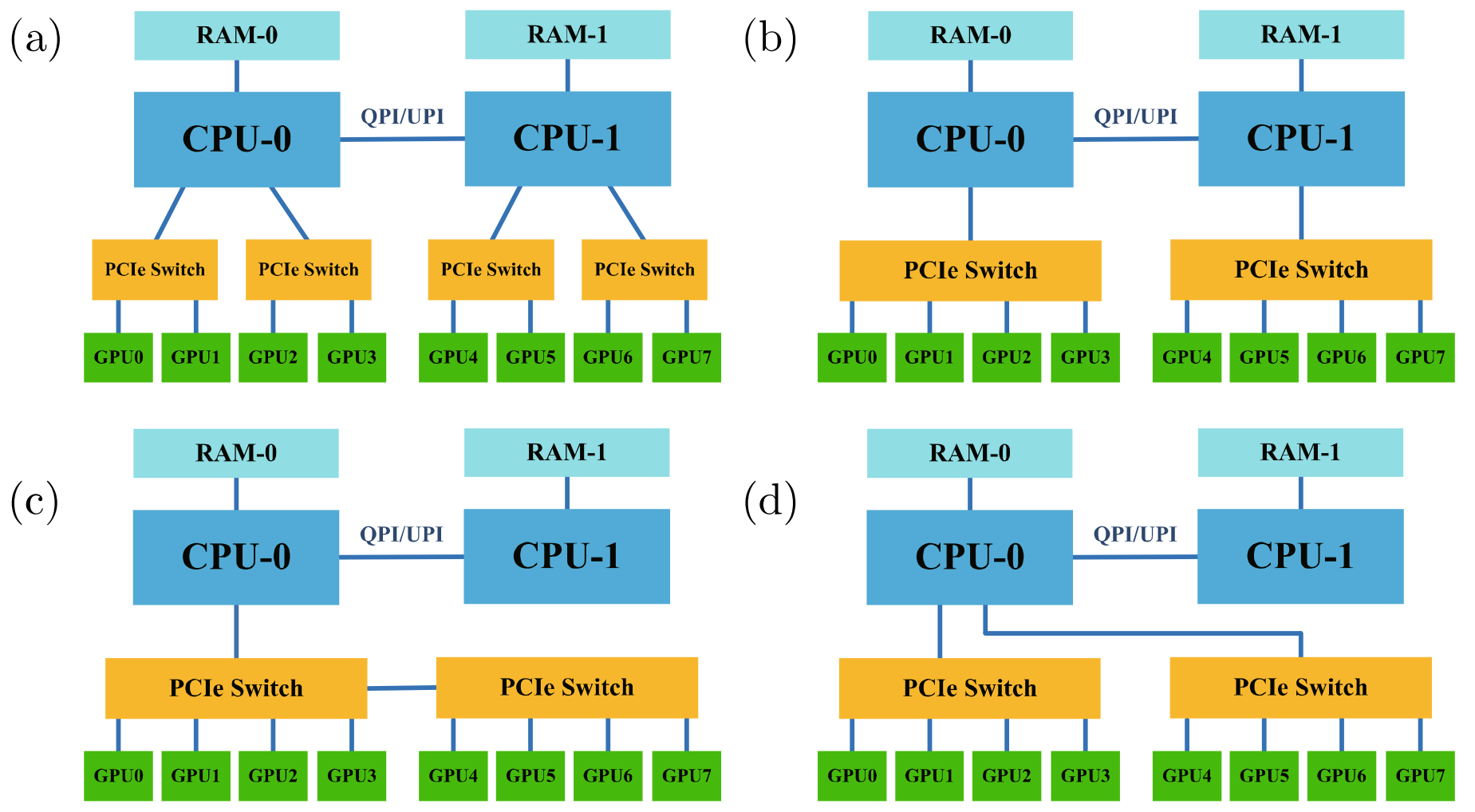

**Fig. 6** Several typical PCIe root architectures of GPU server. For type (**a**), each NUMA node has a PCIe switch attached on them, and there are two GPUs mounted on each PCIe switch. For type (**b**), only one PCIe switch is mounted on each NUMA node, and there are 4 GPUs attached on one PCIe switch. For type (**c**), all GPUs are mounted on one NUMA node by two PCIe switches. For type (**d**), all GPUs are mounted on one NUMA node

Complex PCIe root architecture will affect the efficiency of memory access of CPU-GPU and even process unit (PU) to RAM. Figures 7a and b show the memory access models of GPU to CPU and GPU to GPU. The server has two CPUs interconnected with QPI, and it is a dual root system. In Fig. 7a, a process is running in process core PU0 in CPU-0 and selects GPU0 for acceleration. The data of this process is allocated in the memory (RAM) in the same root with CPU-0 and GPU0. As a result, the process visits its data in RAM with optimal bandwidth, and GPU0 also transfers data to the CPU with optimal bandwidth. This is the ideal situation. However, the PU in which the process is running is not settled and is managed by the system by default. For the pink path in Fig. 7a, the process is running in PU2 in CPU-1 but selects GPU2 as the accelerator. Though it visits data allocated in the closest RAM with optimal bandwidth, GPU2 is not in the same root with CPU-1. Data transfer between CPU and GPU must traverse QPI because of unsupported direct memory access, which is very inefficient. The worse case is the blue path. The data is allocated in RAM close to CPU-0 and GPU1, but the process runs in PU of CPU-1, resulting in poor efficiency of both GPU memory access and RAM access.

PCIe root architecture will also affect GPU to GPU communication or memory access. Figure 7b gives several possible models of GPU to GPU memory access in current GPU server. The red path is the worst case since it traverses QPI, and the topology of GPU2 and GPU3 is usually marked as “SYS”. GPU4 and GPU5 are attached to the same IOH chip for brown path, which provides better communication efficiency than “SYS”. The topology is usually called “PHB”. The purple path connects two GPUs in the same PCIe switch, which provides better efficiency than “PHB”, and its topology is marked as “PIX”. The green path is the best. GPUs interconnect each other directly with Nvlink, which is the fast way of GPU to GPU communication so far, labeled as “NV”.

To exploit the optimal efficiency of the hardware system, users should know the topology of the hardware system. The Hardware locality (HWLOC) software package provides an approach to gather hierarchical topology information about modern increasingly complex parallel computing platforms, including NUMA (Non-Unified Memory Access) memory node, shared cache, cores, and multithreading, as well as I/O devices such as GPUs, InfiniBand HCAs. With the topology information, users can optimize their programs to obtain the best performance. It should be noted that the optimizations of CPU-GPU communication and GPU-GPU communication are interdependent, which will be introduced next.

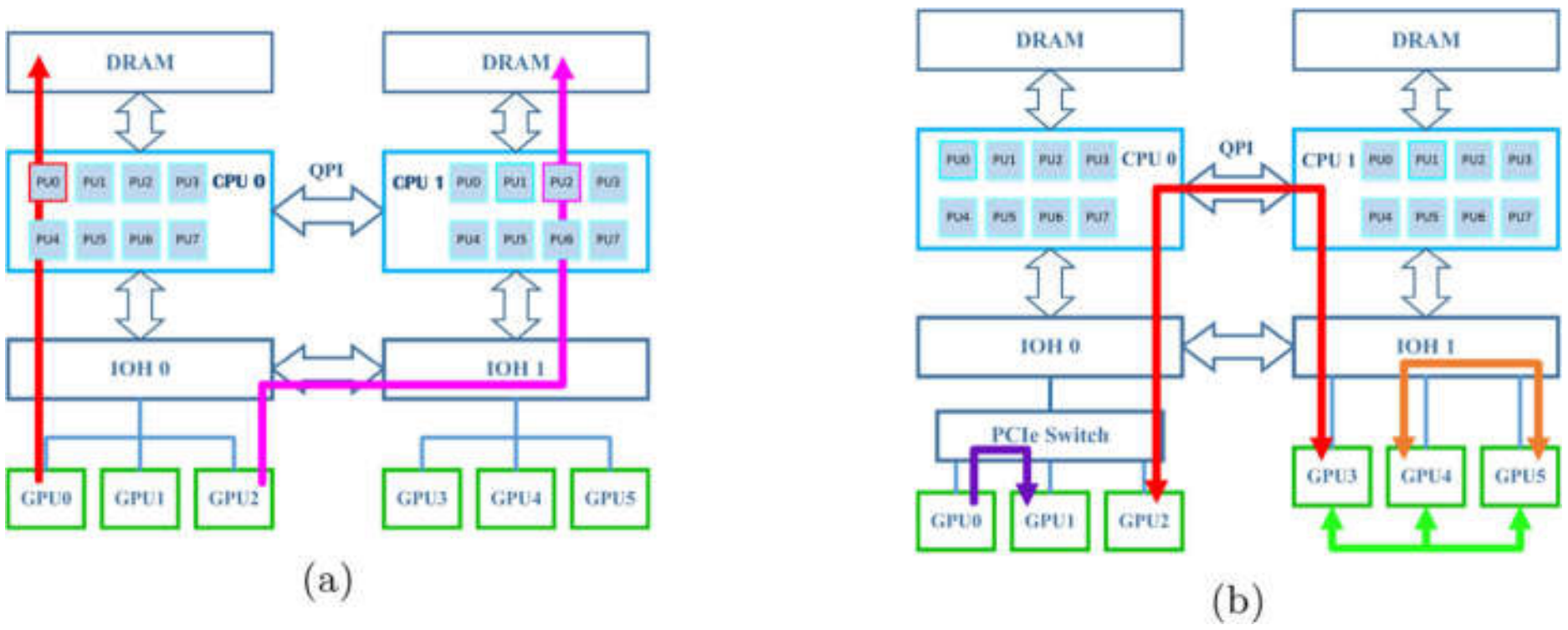

**Fig. 7** CPU-GPU and GPU-GPU memory access models. In (**a**), red path is the optimal one, while pink path is inferior because CPU to GPU memory access across NUMA nodes is worse than local access. In (**b**), if GPUs support peer-to-peer (P2P) communication technology, green path indicates that two GPUs communicate by Nvlink, which is the fast path. Purple path indicates that two GPUs communicate by PCIe switch (PLX), which is the most efficient communication way without Nvlink. Yellow path means that two GPUs communicate by host (CPU). Red path is the worst one because P2P is not supported across NUMA nodes, and memory access between two GPUs in this way must be transferred by both CPUs

#### *4.3.1 CPU-GPU communication optimization*

To optimize the performance of memory access between CPU and GPU, the only way is to get them "closer". The word "close" is an iconic description of two devices that have larger data transfer bandwidth. As aforementioned, the best strategy is to make the RAM, PU, and GPU be utilized by a process located in the same root. There are many ways to bind a process to a specified PU, including the process binding technique of HWLOC, APIs of NVIDIA Management Library (NVML), and process binding technique in Linux system.

Many commercial and open-source software packages provide memory binding. MVAPICH allows data allocation when calling "MPI_Init()", and OpenMPI allocates it after calling "MPI_Init()". The problem is that the size of memory space to allocate before the program calling "MPI_Init()" and reading input files and grids are unknown. We have to run MPI initialization to get the process ID to determine which grid files to read. What is more, to optimize the GPU-GPU communication, we should know the communication data load of each process. Then we can decide which GPU to be selected for acceleration. After the GPU of a process is chosen, we can bind the process to the CPU that the selected GPU is attached to. Finally, we allocate memory and bind it to the CPU. The steps of the procedure are briefly outlined:

- Run the program first until the processes get its GPU ID. When the solver runs in parallel, each process gets its ID and reads its input files accordingly. The process calculates the communication data load according to block connection information. Based on the communication data load and the topology of GPUs of the local machine, the GPU-GPU communication optimization will select the optimal GPU for the process.
- Find the NUMA node (CPU) near the given GPU ID and bind the current process or thread to it. Here we utilize the HWLOC software package to do it. Firstly, set the cpuset of given GPU ID with API function *hwloc_cudart_get_device_cpuset*(). Secondly, bind current process or thread to this cpuset with *hwloc_set_cpubind*(). Finally, get the NUMA node of this cpuset with *hwloc_get_obj_covering_cpuset*().
- Allocate memories in RAM for data of the process. If there is allocated memory of data, free it or migrate it to the CPU selected above. The API function *hwloc_alloc_membind*() in HWLOC is utilized to allocate memory for data in the RAM attached to the selected CPU. This API function will allocate memory in a given size in RAM of the local NUNA node and return the address of this memory space. For the memory of data that has been allocated before GPU ID is got, we can migrate it to the right NUMA node with API function *hwloc_set_area_membind*() when the address and NUMA node ID are provided.

#### *4.3.2 GPU-GPU communication optimization*

GPU-GPU communication is relatively easier to perform, which is efficient if GPUDirect for Peer-to-Peer (P2P) technology developed by Nvidia is applied. There are many Nvidia

GPUs that support GPUDirect P2P, such as Tesla series and Quadro series GPUs. GPUDirect P2P technique includes P2P memory access and P2P transfer and synchronization, and it is an optimization technique of GPU-GPU communication in the same system. With it, buffers between memories of two GPUs in the same computer can be copied directly through PCIe lane or Nvlink. The efficiencies of GPU-GPU communication by use of the GPUDirect P2P technique vary with the path the peers are connected with. Nvlink offers maximum bandwidth. Thus it is the most efficient path. Nvidia has developed a new Nvlink technology, which enables the interconnection of any pair of GPUs in a single system that has multiple GPUs. This strategy works well. However, this is a highly customized product that needs specified motherboards and only supports Tesla P100 SXM and Tesla V100 SXM2. There are many GPU servers that do not support Nvlink, and GPUs are still installed in PCIe slots. Therefore, the GPU-GPU communication optimization is directed to those GPUs communicating through PCIe.

To optimize the GPU-GPU communication through PCIe, we need the topology of the local system and the sizes of buffers to be exchanged among processes located in this server. Then we can distribute process groups that include dense communication load to GPU groups whose communication is more efficient, thus improving global communication efficiency. The details of optimization are introduced below:

- The solver runs in parallel, and each process reads its input file and gets its communication buffer sizes to other processes. Each process may communicate with zero or one process or multiple processes, which may be distributed in one or multiple servers.
- Gather processes located in the same server and let them know each other. MPI software package offers APIs to achieve it. Firstly, a new communicator in local server can be created with API function *MPI_Comm_split_type*() with split type "*MPI_COMM_TYPE_SHARED*", to better distinguish below we call it "*MPI_SHARED_LOCAL*". The size of a new communicator and local rank (or ID) in the new communicator of the current global process can be obtained with APIs *MPI_Comm_size* and *MPI_Comm_rank* respectively. Secondly, get the groups of the global communicator "*MPI_COMM_WORLD*" and local communicator "*MPI_SHARED_LOCAL*". Thirdly, gather the processes that the current process communicates with, and translate them into local communicators. A global process ID that is not located in the local communicator will be marked as "*MPI_UNDEFINED*" so that we filter out processes to communicate within the local server.
- Each process in local communicator shares its information with all other local processes. The information here is the sizes of communication buffers to target processes in local communicators. Then each local process knows the local communication network and buffers size of each connecting line. This procedure can be easy to implement with the MPI API function *MPI_Allgather*().
- Each local process inquires the topology of GPU pairs in the local server. We can obtain the connection type of two GPUs by use of NVML API function *nvmlDeviceGetTopologyCommonAncestor*(*device1, device2,* &*topo_type*) according to the returned variable *topo_type*. The connection type of two GPUs in the same system should be "NV", "PIX", "PXB", "PHB", or "SYS", which range in the order of efficiency decrease. We will not consider "NV" as mentioned above. Then we organize GPUs in groups. Firstly, each CPU (NUMA node) gathers GPUs that are attached to it. Any two GPUs in this group communicate with each other with efficiency not worse than "PHB". Secondly, in each CPU, gather the GPUs in the branch in which any two GPUs communicate with each other with efficiency not worse than "PXB". Thirdly, in each "PXB", gather the GPUs that are in the same PCIe switch. GPUs in a single PCIe switch have connection type "PIX" with each other, and there may be multiple "PIX" groups attached to a GPU.

- Filter out the busy GPUs in the groups above and get groups that contain idle GPUs only. The utilization of GPU can be inquired with NVML API function *nvmlDeviceGetUtilizationRates*().
- Partition local processes according to the number of GPUs of each CPU. If all local processes are less than the GPUs of any CPU, choose the CPU that can hold all local processes. If no single CPU can hold all processes, divide local processes into partitions according to the number of GPUs in each CPU, and make sure that these partitions have minimal total communication data with each other.
- In each CPU that gets a partition, the processes are divided into partitions for "PXB" groups in the fashion above. The "PIX" groups should be dealt with in a similar fashion.

With the above procedures, each process in the local server is binded to the GPU, making it communicate efficiently with other GPUs attached to its target communicating processes. It is worth mentioning that, in modern GPU clusters, servers usually interconnect with the InfiniBand network. With GPUDirect Remote Direct Memory Access (RDMA) technology, buffers can be directly sent from the GPU memory to a network adapter without staging through host memory. In this case, the network adapter should be included during optimization above, and we can simply regard it as a communication target-like process.

### 4.4 Memory utilization

How to utilize memory of GPU is one of the key points in programming GPU solver. Strategies of memory utilization of GPU vary from code to code in different fields according to algorithms applied and computing scales. In the early stage of GPU computing, GPUs were mainly designed for image and video processing which usually involves simple algorithms like matrix operations. Performing simple algorithms on a bunch of single datasets would not occupy much memory of GPU. Hence GPUs several years ago were equipped with very low memory capacity, which is usually not higher than 4GB. For scientific computing that involves large-scale computation, memory occupation is a challenge for CPU; thus, it is hard to hold all data in GPU memory. Due to the limitation of memory capacity, GPU usually works as the local accelerator, where most of the data of program are fixed in the memory of CPU, and a part of data is uploaded onto GPU for computation acceleration, and then the result is downloaded back to CPU. This strategy enables the solver to deal with much more data with limited memory of GPU. Data transfer between CPU and GPU through PCIe lane is time-consuming, which must be compensated in GPU computing in order to get acceleration. However, with the performance of GPU improving greatly, massive and reduplicative data transfer gradually becomes the bottleneck of GPU computing.

The bandwidth of PCIe has not increased significantly over the past few years, but the memory capacities of GPUs have increased a lot. In particular, some professional GPUs are designed for scientific computing, equipped with memory up to 32GB. Moreover, the current GPU server contains more GPUs, making a single server equipped with powerful computational capacity. Hence, the strategy of memory utilization should be adjusted to adapt to current GPU computing. We will introduce the detail of memory utilization of our solver in the following.

There are two basic and vital principles of memory utilization in our solver. On the one hand, allocate fixed permanent memory space for the key dataset. On the other hand, avoid reduplicative memory allocation and deallocation by allocating a piece of permanent memory space that is reasonably large enough for potential public usage.

In the whole procedure of solving Eq. (2), we select several fundamental and dominant variables from all variables. A variable is fundamental when it is used throughout the whole computation, and many other variables can be obtained with them. The metrics $(\xi, \eta, \zeta)_{x,y,z}$ and Jacobian $J$ are utilized all over the computation, and they are constants if the grid is stationary, so they are fundamental variables. The conservative flow variable $\widetilde{U}$ in Eq. (3) is fundamental as well because it plays a key role in time advancement in calculating the interpolated variables in WENO interpolation. Moreover, it is also the source to compute primitive variable $Qv = (\rho, u, v, w, p, T)$. For convenience in utilization, we will not include the Jacobian $J$ in it and only store $Q = (\rho, \rho u, \rho v, \rho w, e)$. Although $Qv$ can be obtained from $Q$, it is always used directly during the whole computation. So we have to sacrifice memory to avoid unnecessary computation time. For viscous flow computations, molecular viscosity $\mu$ is unavoidable, which is used in computing viscous flux and local time steps. In time advancement with the Runge-Kutta scheme, as shown in Eq. (20), $Q^n$ is a variable that should be stored. For time advancement in the local time step, the time step $\Delta t$ at each grid point should also be stored. $R(Q)$ represents the term of the right-hand side and contains the contribution of derivatives of inviscid flux and viscous flux, and is used in time advancement and residual computation. Thus $R(Q)$ should be stored during the whole computation. In the following, we denote the size of the original grid of the current process with $N$ and the size of the grid containing ghost cells with $M$. According to the above analysis, we need to allocate memory for fundamental and dominant variables with $32 * M$ elements.

During computations of derivatives of inviscid and viscous fluxes, there are many temporary variables at each grid point. According to numerical methods applied in inviscid terms, we need memory space to store the physical flux $f_i$ and numerical flux $f_{i\pm\frac{1}{2}}$, as shown in Eq. (19). The total size of temporary variables required is $10 * M$. In the viscous term, the derivatives of $(u, v, w, T)$ with respect to $(x, y, z)$ are widely used. Moreover, viscous flux has to be stored according to our kernel mentioned in the next section. In total, we need to allocate memory space for temporary variables in the viscous term with the size of $17 * M$. However, the inviscid term and viscous term are computed successively, and there is no need to allocate memory space for temporary variables in the inviscid term and viscous term, respectively. Thus we allocate memory space with a size of $17 * M$ elements, which is enough for current public usage and potential usage elsewhere.

In short, we need permanent memory space with a size of $49 * M$ elements. The variables mentioned above are all in float type. In our solver, arrays in integer type are widely applied, such as in blocks communication and physical boundary conditions. We allo-cate memory for arrays for public usage in the way mentioned above too. Our solver can avoid unnecessary data transfers between CPU and GPU. Meanwhile, we save the GPU memory up to the hilt and increase the maximum grid size the solver can deal with. The maximum grid capacities of our solver in different single GPU are shown in Table 3. The data is stored in double precision float type, and the result is satisfying for a single GPU. The computation speeds in full load are also well accepted in general applications because more grids will lead to unacceptable computation speed. Therefore, the memory strategy mentioned above is well matched with our programming algorithm strategy.

**Table 3** Maximum grid capacity of ***HiResX*** solver in different GPUs

| GPU | Memory capacity (GB) | Maximum grid capacity (million) |
|---|---|---|
| RTX 2080 Ti | 11 | 25 |
| Tesla P100 | 16 | 35 |
| Titan V | 12 | 27 |

### 4.5 CUDA kernels

The efficiency of GPU computing relies heavily on algorithms and programming skills. To fully exploit the parallel capability of GPU, one needs a deep understanding of the hardware of GPU and its execution mechanism. The GPU architecture consists of a scalable array of Streaming Multiprocessors (SM). Figure 8 shows the structure of GPU in software and hardware layers. Each SM in a GPU is designed to allow concurrent execution of hundreds of threads. Generally, there are multiple SMs per GPU, so it is possible to have thousands of threads executing concurrently on a single GPU. In CUDA, all the threads are organized in a group called thread block. A thread block can only be scheduled on one SM and remains on that SM until execution finishes. Hence, how to utilize threads is one of the primary things in kernel design.

In the CUDA memory model, many types of memory are programmable to the user, as shown in Fig. 8. Global memory is the largest, highest-latency, and most commonly used memory on a GPU, and it is accessible to any SM throughout the lifetime of the application. Registers are the fastest memory space on a GPU and are partitioned among active warps in an SM, so register variables are private to each thread. Registers store automatic variables declared in a kernel without any other type qualifiers. Arrays declared in a kernel with a constant that can be determined at compile time are stored in registers. Local memory in nature is a part of memory that resides in global memory, which is used to store variables in a kernel that are eligible for registers. However, it cannot fit into the register space allocated for that kernel, including arrays referenced with indices that cannot be determined when code is compiled. Shared memory is a type of on-chip memory space with much higher bandwidth and lower latency than local or global memory. Registers and shared memory are scarce resources. These limited resources impose a strict restriction on the number of active warps in an SM, thus affecting the parallel performance in an SM. Therefore, how to utilize these types of memory is another thing that is vital to performance in kernel design.

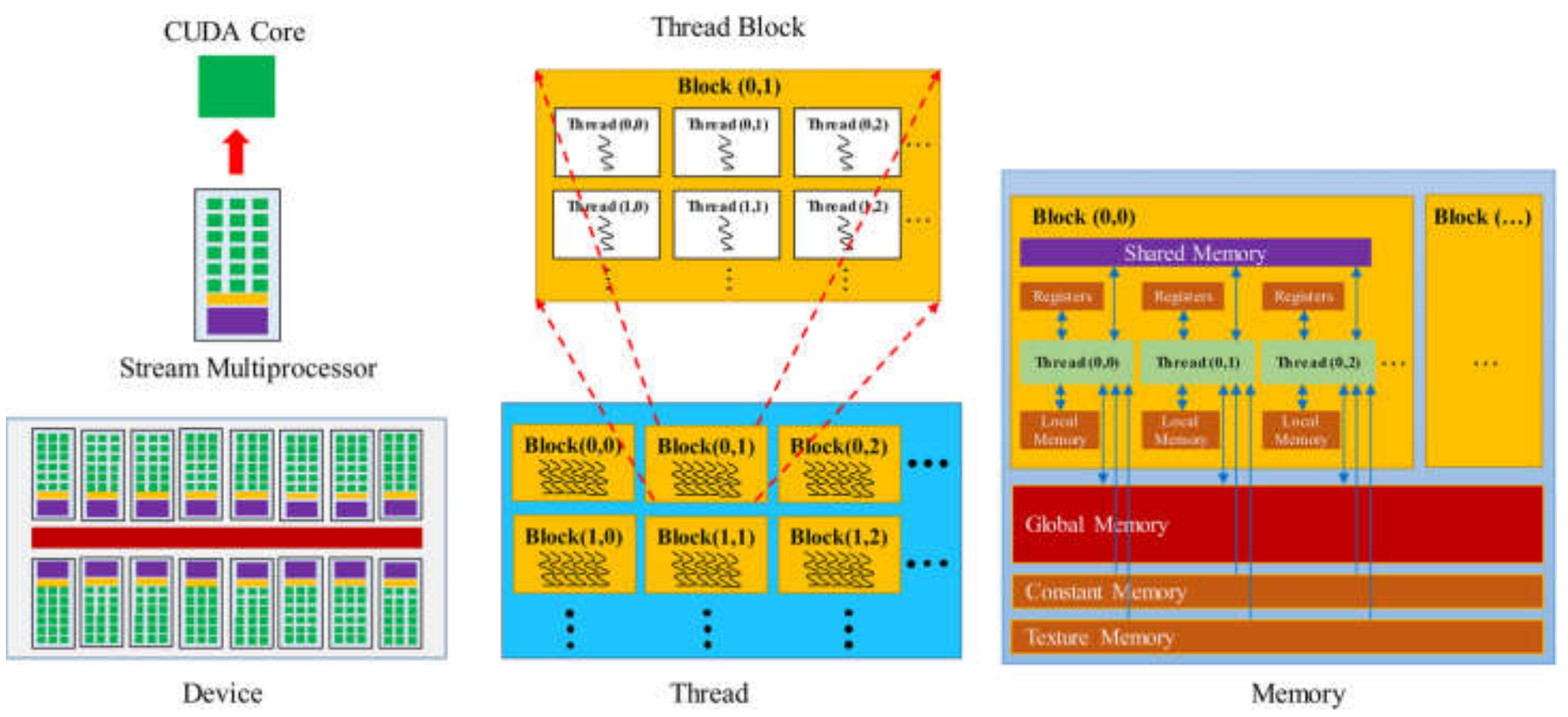

**Fig. 8** Organization of CUDA threads for NVIDIA's GPU and its mapping to hardware structure

In GPU computing, the CPU is called host, and GPU is called device. The kernel function is a piece of code that runs on the device, and the design of kernel functions is the core of GPU computing. In current computation, most of the computation in solving Eq. (2) is spent on evaluation of inviscid term and viscous term. Evaluating inviscid term and viscous term involves much more complex operations than other sections of computation. Therefore we will give more details of them in the following.

#### *4.5.1 Inviscid fluxes*

We consider a three dimensional grid block with sizes of $(NI, NJ, NK)$ in $I, J, K$ directions. In CPU source code, derivatives of convective fluxes with respect to $\xi, \eta, \zeta$ directions are evaluated in sequence, and in each direction we perform the procedure from Eq. (10) to Eq. (19). Take the derivative in $I$ direction as an example, in the $JK$ face of grid block, $NJ * NK$ one-dimensional problems in $I$ direction are solved. For each pipeline in $I$ direction, conservative variables at half-point or edge location $(i + 1/2)$ are evaluated with fifth order WENO interpolation as is shown in Eq. (13) to Eq. (17). To minimize oscillation, interpolation is performed in characteristic space. There are two interpolated values, $u^+_{i+1/2}$ and $u^-_{i+1/2}$, at each half-point location, which are approximated with values of left-biased point stencil $(i-2, i-1, i, i+1, i+2)$ and right-biased point stencil $(i-1, i, i+1, i+2, i+3)$ respectively. Then at each half point, Steger-Warming flux is used to construct numerical flux $f_{i+1/2}$. The physical flux at the grid point can be simply obtained with Eq. (4). Finally, the derivative at each point can be approximated with Eq. (19). Derivatives in the other directions can be done in the same way, and the results are accumulated during the loop of direction to construct the term $R(Q)$ in Eq. (20) for Runge-Kutta integration.

All above computation is done by a single process when no multi-threads technique like OpenMP is applied. The code running on a GPU device should do all the work as it does on the CPU. The structure of GPU and its execution mechanism have been introduced earlier, and it is noted that we can call tens of thousands of threads to complete certain computations. To port a code onto GPU, the first work to do is how many threads to call and arrange their works. In our early version of our GPU code, a thread is assigned to complete a 1D problem, so the number of threads to be called is the size of the sweeping face of grid block, $NJ * NK$ for JK face. However, the acceleration performance is poor. This problem is because the number of active threads is severely restricted by the high occupation of the register in the kernel function. In addition, the number of threads to be called is not large enough and is limited by the sweeping faces of grid block, which usually vary with sweeping directions.

Two improvement measures are introduced to solve this problem. The first improvement measure is that atomic level operation is applied. Other than assigning each thread to complete a 1D problem of the whole grid line, we assign each thread to complete the work at only a grid point. It is a so-called "atomic level operation". Thus the number of threads to be called increases up to the number of total grid points, and it does not vary with sweeping directions of derivatives. The CUDA kernel is therefore configured as

$$
\begin{aligned}
&int \quad blocksize \quad = \quad BLOCK_SIZE_INVISC;\\
&int \quad nblock \quad = \quad ((\,NI * NJ * NK \; - \; 1)\;/\;blocksize\,) + 1;\\
&cuda_afweno5 <<< \; nblock, \quad blocksize \; >>> (...);
\end{aligned}
$$

Generally, optimizing the workloads to fit within the boundaries of a warp, a group of 32 threads, will lead to more efficient utilization of GPU compute resources. Thus, the value of *BLOCK_SIZE_INVISC* should be divided evenly by 32. The improper value that is too small or large may cause performance reduction. In practical applications, 256 is usually proved to get satisfying performance. But the best value should be determined by testing.

The second improvement measure is the utilization of the kernel decomposition technique. By decomposing a big kernel function into several small ones, the number of registers needed in kernel function reduces, and the number of active threads increases. Figure 9 shows the pseudo-code of computation of derivatives of convective flux. The procedure of evaluating derivatives is split into three parts, and the computation of each part is done with a kernel function. In kernel **cuda_afweno_split_flux(...)**, numerical flux $f_{i+1/2}$ is computed at each half-point by a thread, and the result is written into public memory space. After the kernel finishes, numerical fluxes at all half-point locations in that direction are already known. Then, the second kernel function **cuda_phys_flux(...)** computes physical flux $f_i$ at each point by a thread, and the result is written into the public array in global memory too, and at the end of it, physical fluxes are all known. At last, in the third kernel function **cuda_cflux_df(...)**, numerical flux and physical flux are obtained from global memory to compute derivatives at each point by a thread, and the result is stored in array $R(Q)$ in global memory. Theoretically, the second kernel can be merged with the first kernel. But the merged kernel function occupies too many registers, and the number of active warps is strictly restricted due to limited registers. By separating the independent part and computing it with a small kernel, more warps are active, thus improving parallel performance. However, we do not divide the kernel **cuda_afweno_split_flux(...)** into two parts to calculate interpolation of conservative variables and numerical fluxes as is done in [6], because delivering data from one kernel to another kernel involves access to global memory, which brings high latency. Moreover, these two parts are complex as well, and the kernels still occupy too many registers. So the number of active threads does not increase. Our test shows that the time saved by increasing active threads does not compensate for the latency of global memory access. It is worth mentioning that metrics and Jacobian at half-point, which are needed to compute numerical flux, are neither stored in global memory nor transferred from CPU. If we save them in global memory permanently, more memory is required for the solver to deal with certain grid sizes. Although the public memory space is large enough to hold them, we do not transfer them from the CPU either because our test suggests that computing is

```
loop idir→  direction in I, J, K
        __global__ cuda_afweno_split_flux(...)
           – calculate left matrix L at i+½
           – map U at point (i−2, i−1, i, i+1, i+2) into characteristic space and get
             characteristic variables V
           – perform WENO interpolation to get V+ at i+½
             −− evaluate V with three substencils, Eqn (13)
             −− evaluate smoothness indicators of each substencils, Eqn (16)
             −− evaluate nonlinear weights, Eqn (14)
             −− perform nonlinear combination, Eqn (12)
           – map U at point (i−1, i, i+1, i+2, i+3) into characteristic space and get
             characteristic variables V
           – perform WENO interpolation to get V− at i+½
             −− evaluate V with three substencils, Eqn (13)
             −− evaluate smoothness indicators of each substencils, Eqn (16)
             −− evaluate nonlinear weights, Eqn (14)
             −− perform nonlinear combination, Eqn (12)
           – calculate right matrix R at i+½
           – mapping characteristic variables back to physical space
           – calculate numerical flux f(U+) and f(U−) at i+½  with Riemann solver
           – compute f(U+,U−) = f(U+) + f(U−) at i+½
        __global__ cuda_phys_flux(...)
           – compute F(U), with Eqn (3)
        __global__ cuda_cflux_df(...)
           – compute derivative with Eqn (18)
end loop direction in I, J, K
```

**Fig. 9** Code structure of derivative of inviscid flux computed with original AFWENO

much faster than transferring. Thus they are interpolated from point values directly when they are needed.

We have known that in GPU, as processor-to-memory latency increases, the capacity of that memory increases. Global memory is the largest, highest-latency memory space on a GPU. So frequent access to global memory will increase the execution time of a kernel. Therefore, reducing access to global memory provides the possibility of optimization during programming and numerical algorithm design. In the current work, we also propose an improved WENO interpolation, which reduces access to global memory and computation price. Figure 10 shows the schematic diagrams of original WENO interpolation and improved WENO interpolation. In original method, computing numerical flux at $i+1/2$ needs points stencil $(i-2, i-1, i, i+1, i+2, i+3)$, and performs two complete WENO interpolations from Eq. (13) to Eq. (17). The two WENO interpolations share the same characteristic space located at $i+1/2$. In CUDA kernel function **cuda_afweno_split_flux(...)** mentioned above, for the Steger-Warming splitting flux method, $f^+_{i+1/2}$ and $f^-_{i+1/2}$ are obtained simultaneously in the same thread, thus numerical flux is obtained immediately with $f_{i+1/2} = f^+_{i+1/2} + f^-_{i+1/2}$. In our improved method, we build the characteristic space at $i$, and we utilize the points stencil $(i-2, i-1, i, i+1, i+2)$ to perform WENO interpolation to get $u^+_{i-1/2}$ and $u^-_{i+1/2}$. We do not need to perform

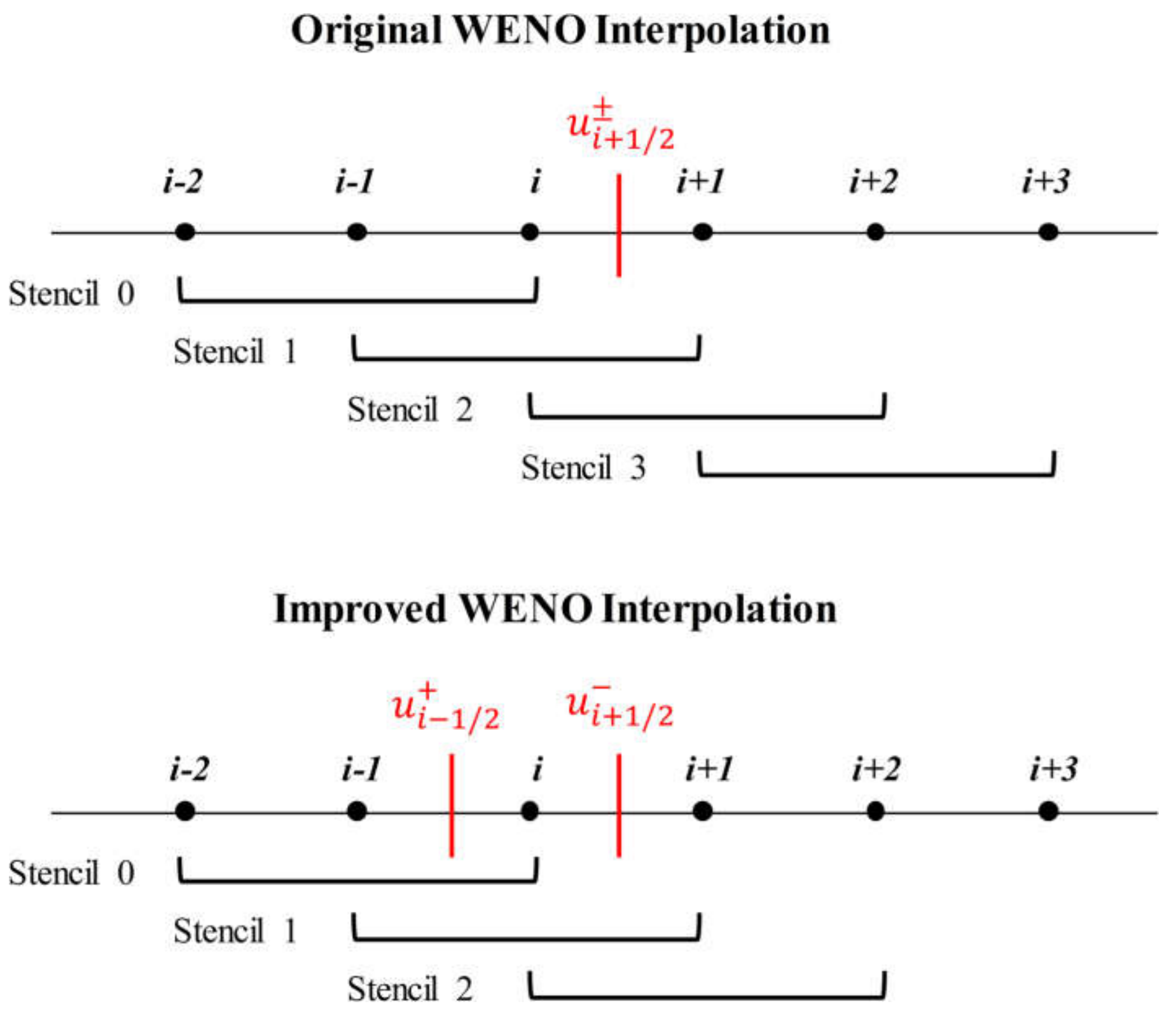

**Fig. 10** Comparison of WENO interpolations. In original WENO interpolation, the characteristic space is built at $i+1/2$, and $u^{\pm}_{i+1/2}$ are computed simultaneously in a loop. In modified WENO interpolation, the characteristic space is built at $i$, and $u^{-}_{i+1/2}$ and $u^{+}_{i-1/2}$ are computed simultaneously in a loop. In modified WENO interpolation, point stencils are the same, so the characteristic variables and smoothness indicators are shared, thus it reduces calculation and global memory access

two complete independent WENO interpolations because the two WENO interpolations share the characteristic variables and the smoothness indicators. Since we can not get $f^{+}_{i+1/2}$ and $f^{-}_{i+1/2}$ in the same thread, thus we have to compute numerical flux $f_{i+1/2}$ with a new kernel function **cuda_sum_split_flux(...)**. The procedure is shown in Fig. 11.

#### *4.5.2 Viscous fluxes*

Evaluation of viscous terms is slightly easier than convective terms. According to Eqs. (5) and (6), we have to compute $(u, v, w, T)_{x,y,z}$ before computing viscous fluxes. Here, we approximate the derivative of viscous flux with fluxes defined at point locations, so we need to compute $(u, v, w, T)_{x,y,z}$ once and share them during derivatives in all directions. Computing $(u, v, w, T)_{x,y,z}$ is easy by the chain rule. In the general curvilinear coordinate system, derivatives with respect to $(x, y, z)$ can be obtained with chain rule as follows

$$\begin{aligned}
\frac{\partial}{\partial x} &= \xi_x \frac{\partial}{\partial \xi} + \eta_x \frac{\partial}{\partial \eta} + \zeta_x \frac{\partial}{\partial \zeta} \\
\frac{\partial}{\partial y} &= \xi_y \frac{\partial}{\partial \xi} + \eta_y \frac{\partial}{\partial \eta} + \zeta_y \frac{\partial}{\partial \zeta} \\
\frac{\partial}{\partial z} &= \xi_z \frac{\partial}{\partial \xi} + \eta_z \frac{\partial}{\partial \eta} + \zeta_z \frac{\partial}{\partial \zeta}
\end{aligned} \tag{21}$$

```
loop idir→ direction in I, J, K
        __global__ cuda_afweno_split_flux(...)
          − calculate left matrix L at i
          − map U at point (i−2, i−1, i, i+1, i+2) into characteristic space and get
            characteristic variables V
          − perform WENO interpolation to get V+ at i−1/2
            −− evaluate V with three substencils, Eqn (13)
            −− evaluate smoothness indicators of each substencils, Eqn (16)
            −− evaluate nonlinear weights, Eqn (14)
            −− perform nonlinear combination, Eqn (12)
          − perform WENO interpolation to get V− at i+1/2
            −− evaluate V with three substencils, Eqn (13)
            −− evaluate nonlinear weights, Eqn (14)
            −− perform nonlinear combination, Eqn (12)
          − calculate right matrix R at i
          − mapping characteristic variables back to physical space
          − calculate numerical flux f(U+) at i−1/2  and f(U−) at i+1/2  with Riemann solver
        __global__ cuda_sum_split_flux(...)
          − compute f(U+,U−) = f(U+) + f(U−)
        __global__ cuda_phys_flux(...)
          − compute F(U), with Eqn (3)
        __global__ cuda_cflux_df(...)
          − compute derivative with Eqn (18)
end loop direction in I, J, K
```

**Fig. 11** Code structure of derivative of inviscid flux computed with modified AFWENO

$(u, v, w, T)_{\xi,\eta,\zeta}$ can be approximated directly with sixth-order central scheme. In CPU code, the computation strategy of the viscous term is similar to the inviscid term. After $(u, v, w, T)_{x,y,z}$ are computed, viscous flux and its derivative can be obtained in each direction.

We port the code to GPU, similar to the convective term. Figure 12 shows the procedure of evaluation of viscous term. There are two parts and each part involves loops in directions. In the first part, contribution of $(u, v, w, T)_{x,y,z}$ in each direction is computed, accumulated, and stored in public array in global memory. For example, in $\xi$ direction, $\xi_x \frac{\partial(u,v,w,T)}{\partial\xi}$, $\xi_y \frac{\partial(u,v,w,T)}{\partial\xi}$ and $\xi_z \frac{\partial(u,v,w,T)}{\partial\xi}$ are computed. This part is complete by only one

```
loop idir→ direction in I, J, K
        __global__ cuda_vars_derv_xyz(...)
            − calculate derivatives of (u, v, w, T) in (ξ, η, ζ) directions
            − calculate derivatives of (u, v, w, T) with respect to (x, y, z) with chain rule
end loop direction in I, J, K

loop idir→ direction in I, J, K
        __global__ cuda_vflux_flux(...)
          − caculate viscid flux
        __global__ cuda_vflux_df(...)
          − compute derivative of viscid flux sixth order central difference scheme
end loop direction in I, J, K
```

**Fig. 12** Code structure of the derivative of viscous flux summarized by a pseudo code

kernel function **cuda_vars_derv_xyz(...)**. We have two strategies of thread schedule for this kernel. In the first strategy, the kernel is designed to compute contribution in each direction for four variables $(u, v, w, T)$ at one point by one thread. So the CUDA kernel **cuda_vars_derv_xyz(...)** is therefore configured as

$$\begin{aligned}&int \quad blocksize \;=\; BLOCK_SIZE_VISC;\\&int \quad nblock \;=\; ((\,NI * NJ * NK \;-\; 1)\;/\;blocksize\,) + 1;\\&cuda_vars_derv_xyz <<<\; nblock, \quad blocksize \;>>> (...);\end{aligned}$$

In the second strategy, the kernel is designed to compute contribution in each direction for only one variable at one point by one thread. So the number of threads to be called is four times that of the first strategy, and the CUDA kernel **cuda_vars_derv_xyz(...)** is therefore configured as

$$\begin{aligned}&int \quad blocksize \;=\; BLOCK_SIZE_VISC;\\&int \quad nblock \;=\; ((\,NI * NJ * NK * 4 \;-\; 1)\;/\;blocksize\,) + 1;\\&cuda_vars_derv_xyz <<<\; nblock, \quad blocksize \;>>> (...);\end{aligned}$$

Each thread must determine the indexes of the grid point and the variable according to its thread ID. The indexes of grid points are needed because of the one-side scheme for derivatives near the boundaries of the grid block. In the convective term, we do not need to care about derivatives near the boundaries of grid block.

In the second part, two kernels are applied, which are similar to convective terms. In the first kernel, the viscous flux at each grid point is computed by one thread. After the first kernel completes, the second kernel **cuda_vflux_df(...)** is ready to compute the derivative at each grid point by one kernel, and the result is accumulated into $R(Q)$ which has already included the contribution of convective terms.

#### *4.5.3 Kernels in other section of the solver*

We have introduced the kernels of the convective and viscous terms, which account for most total computation, and they are completely performed on GPU. In order to achieve the whole computation performed on GPU, other parts of the solver have to be ported onto GPU as well. Fortunately, these parts involve only simple algorithms and operations, which are independent of the grid. Therefore, we apply the "atomic level operation" when designing the kernels. For the computation of residual, in which summation of all inner grid points is necessary, we utilize "reduced-dimensional operation" to get better efficiency, although there is an atomic operation in CUDA. "Reduced-dimensional operation" here means that, for three-dimensional grid block, CUDA threads are mapped into grid points of the maximum face of grid block, and each thread performs a specified operation like accumulation over points of that grid line. When the "$3D \rightarrow 2D$" procedure is complete, the "$2D \rightarrow 1D$" and "$1D \rightarrow SingleValue$" are performed similarly. The "reduced-dimensional operation" is also applied to find the minimum time step for unsteady simulation controlled with the CFL number.

## 5 Performance result

### 5.1 Speedup varies with grid size

In this section, we present the performance of ***HiResX*** running on a single GPU with different grid sizes. We compare the performance on different GPUs as well. Three different

GPUs are employed for the test: RTX 2080 Ti, Tesla P100, and Titan V. The specs and the characteristics have been introduced before. The maximum grid capability of the solver on these GPUs is also listed in Table 3. Since the memory capability of RTX 2080 Ti GPU is the smallest one, the maximum grid size tested here is bound to 25 million. The CPU in the test is Intel Xeon E5-2680v3. The solver runs on a single CPU core for CPU computing, and all computation is performed on the CPU. For GPU computing, because GPU code can not run independently of CPU, the present solver uses one CPU core and one GPU device, but nearly all the computation is performed on GPU devices. The speedup here is defined as the ratio of CPU computing time to GPU computing for the same case. Global speedups of the optimal GPU version of the code compared to the serial CPU version for grid sizes arranging from one million to 25 million are shown in Fig. 13. We get significant speedups for different GPUs for all grid sizes. Generally, the speedups for all GPUs increase with the grid size when the grid size is smaller than 15 million, but after that, RTX 2080 Ti and Tesla P100 gradually reach their limits. Thus the maximum speedups for RTX 2080 Ti and Tesla P100 are about 640 and 1500, respectively. Titan V has not shown a drop when the grid size increases. Due to the limitation of memory capacity and the grid capacity, the maximum speedup of Titan V is predicted to be about 1950.

The trend of actually measured performance is matched with their ideal DP performances for these three GPUs, though the amplitudes deviate much from ideal DP ratios. According to the specs of these GPUs in Table 1, the bandwidths of RTX2080 Ti and Titan V are close, but the actually measured performance of Titan V is nearly three times as much as that of RTX 2080 Ti, which benefits from more CUDA cores and the higher

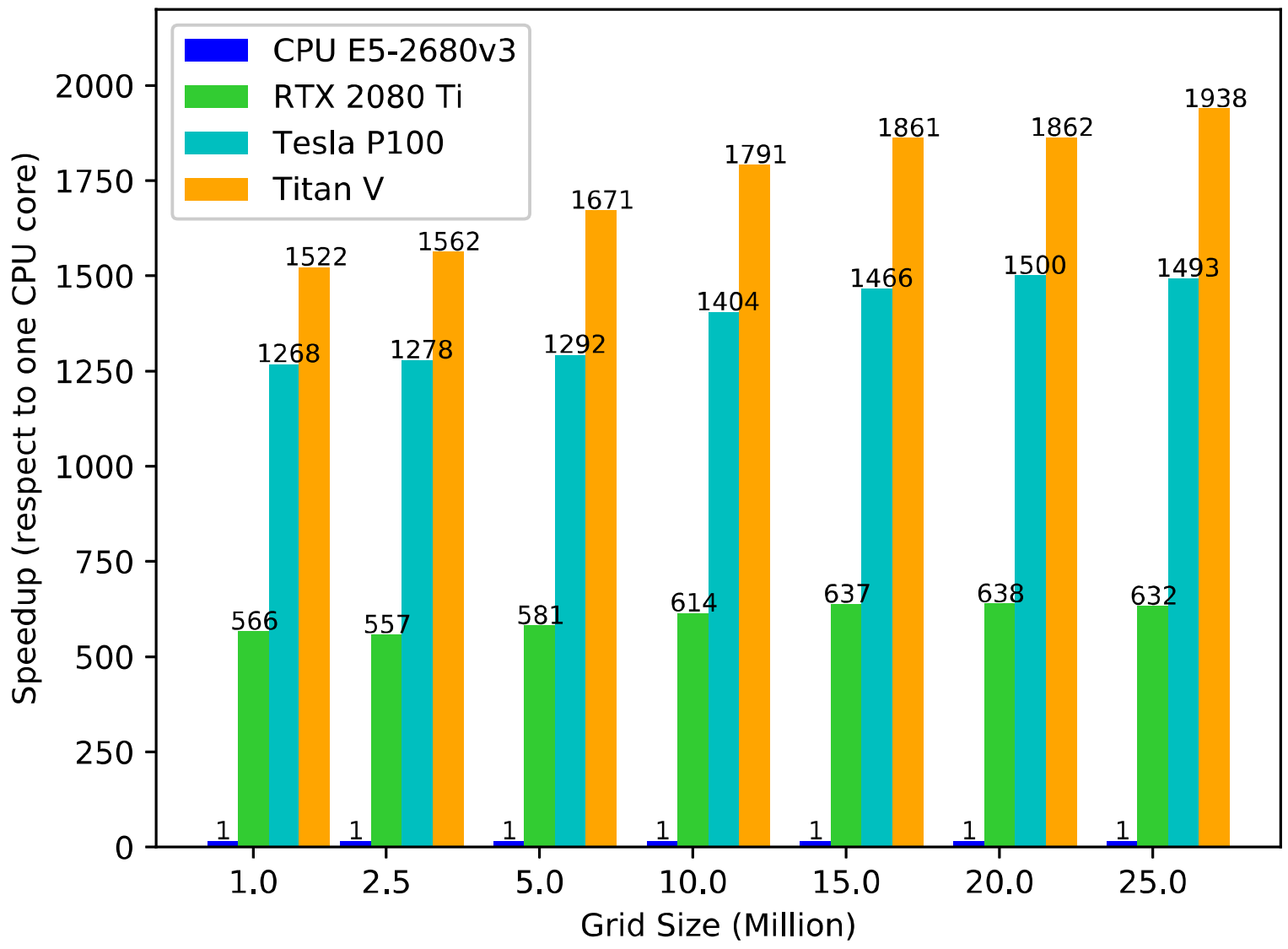

**Fig. 13** Global performance varying with grid sizes. The speedup is defined as the ratio of elapsed time of running on CPU with one core to lapsed time of running with a GPU. The accelerating performance of three GPUs is in agreement with their DP performance (see Table 1). Titan V gets the highest speedup, due to maximum number of CUDA cores and maximum number of double precision processing units. Although RTX 2080 Ti has more CUDA cores than Tesla P100, it has less double precision processing units than Tesla P100. When the grid size increases, speedups of the three GPUs increase in general. But Titan V and Tesla P100 get higher increments. When the grid size is larger than 15 million, Tesla P100 and RTX 2080 Ti almost reach their limits, while there is space for Titan V

ratio of a number of double precision process units to single-precision process units (1:2) in Titan V. Tesla P100 has fewer CUDA cores than RTX 2080 Ti, but has a higher ratio of the number of double precision process units to single-precision process units, 1:2 for Tesla P100 and 1:32 for RTX 2080 Ti, resulting in more double-precision process units. Hence RTX 2080 Ti is mainly bounded to computation rather than memory bandwidth. According to the performance analysis tools of NVIDIA, the large deviation from ideal performance for these three GPUs comes from the relatively low occupancy in convective fluxes and viscous fluxes, which account for the majority of global computation. Low occupancy is caused by the limitation of registers. If there are not enough active threads, the double-precision capability would be excessive, which causes a large deviation from ideal performance. Although we have decomposed the big kernel into smaller ones and cut down the number of registers in the kernel to increase occupancy and number of active threads, high latency of introduction of other types of memory (e.g., global memory) cancels the accelerating effect of higher occupancy. So there is a balance of occupancy and memory efficiency.

### 5.2 Performance of kernels

Global performance relies on the performance of kernels in each part of the solver. Figure 14 shows the performance of several main parts in which all computations are located. As shown in the figure, the part with massive computation gets apparent accelerating, such as evaluation of the inviscid term, viscous term and time step, and procedure of time advance and computation of primitive variable. In these parts, there are a lot of complicated calculations at each grid point. Though evaluation of residual is performed at each point, the kernel for residual computing is not designed to run with "atomic operation", and there is no complicated calculation in the kernel. So the performance of residual evaluation is relatively low. For Runge-Kutta integration in time advancement, there is only a little calculation in this part, but the calculation is performed at each grid point independently, so the "atomic operation" strategy brings apparent acceleration. The speedup of Titan V is double of RTX 2080 Ti because Titan V has more CUDA cores and higher bandwidth. For convective terms, in which a lot of complicated calculation is involved, the speedup of Titan V is relatively high. But the speedup of RTX 2080 Ti is only one-fifth of Titan V, because both the number of double precision process units of RTX 2080 Ti and the number of CUDA core are less than that of Titan V. For viscous terms, in which the computation is second only to the convective term, the accelerating effect is the best. It is worth mentioning that the gap of speedups of both GPUs is much smaller than that of the convective terms. It is the kernel decomposition used to increase the occupancy and active thread that brings the effect. In each kernel in the viscous term, the amount of calculation is medium, so for RTX 2080 Ti the number of double precision process units is not that scarce, but for Titan V it may be excessive. The scenario is similar for the time step.

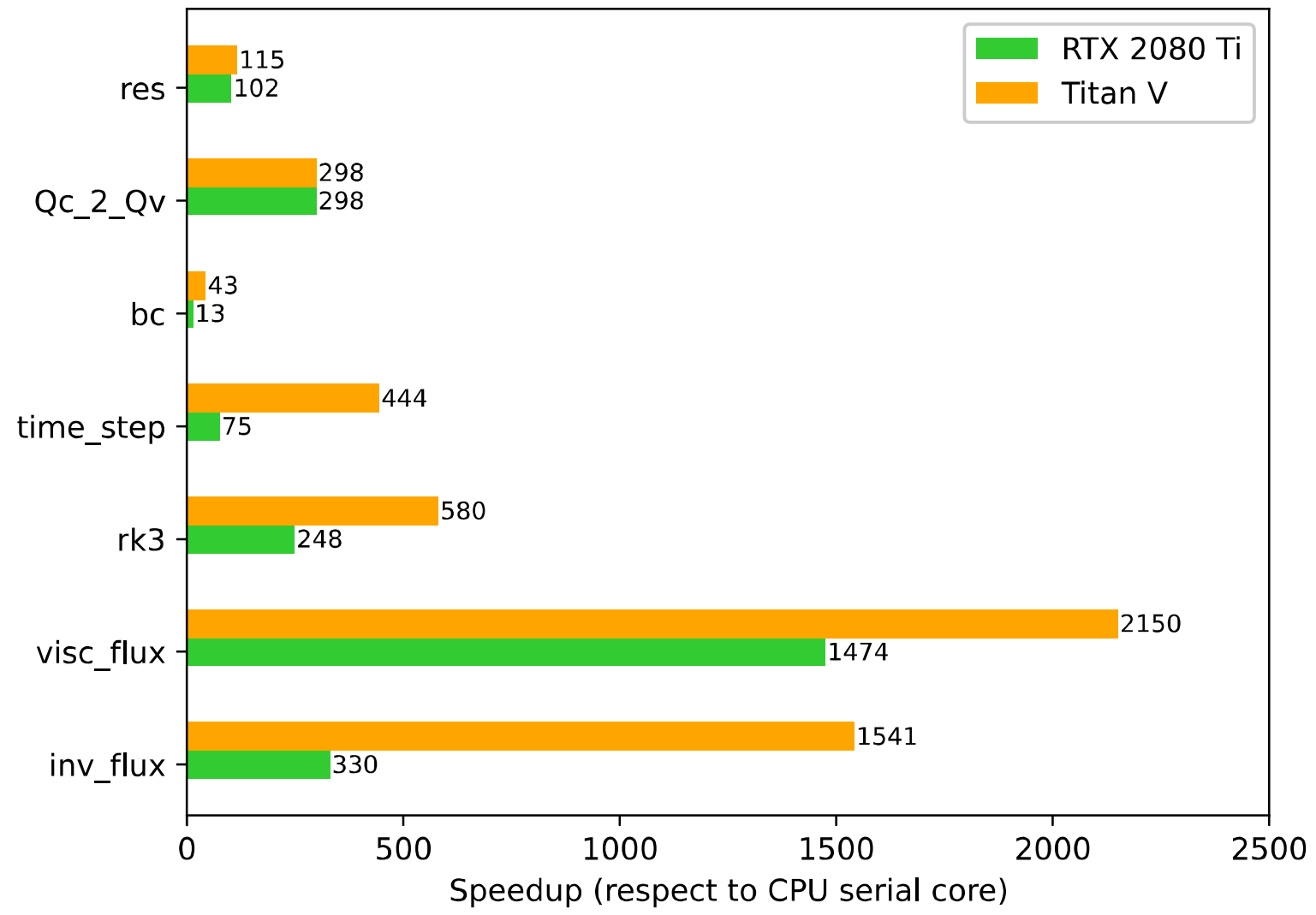

**Fig. 14** Performance of main parts of the solver. Generally, the parts that contain large amounts of computation get apparent acceleration. In the sections of time advance and viscous flux, due to the high occupancies of kernels, both GPUs get higher speedups. For evaluation of time step and inviscid flux, Titan V gets higher speedups than RTX 2080 Ti, because Titan V has more double precision operation processing units, which means higher double precision operation performance, see Table 1

### 5.3 Performance of running with multiple GPUs

In this section, the performance of our solver on multiple GPUs will be investigated. This test aims to show the scalability of our solver when it is running on multiple GPUs. For traditional CPU code, when it runs in parallel, the parallel efficiency is affected by communication among processes and computation loads of processes. In early GPU computing code, besides communication on CPU and computation load, data transfer between CPU and GPU affects parallel efficiency. For modern GPU computing, communication between GPUs is a new factor.

The grid size of all cases in this test is fixed to be 25 million due to the limitation of the memory capacity of RTX2080 Ti. A single block with 25 million grid points is divided nearly equally into several sub-blocks, and each GPU computes one sub-block. Each GPU is driven by a CPU core, so for multiple GPUs cases, the solver runs in parallel with MPI actually. As we have mentioned that for GPU computing, nearly all computations are performed on GPU, so the contribution of CPU to global computation is negligible. The elapsed times of the solver running with the number of GPUs that varies from 1 to 10 are shown in Fig. 15. The scalability gradually deviates from the linear value when the number of GPUs increases, but the scalability efficiency is still larger than 75% when 10 GPUs are utilized. We observe that the performance of scalability of Titan V drops faster than RTX 2080 Ti. However, it is not the parallel efficiency of CPU that causes performance to drop, but the fact that when grid size decreases, the speedup of GPU drops as well. As illustrated in Fig. 13, the variation of speedup of Titan V is greater than that of RTX 2080 Ti. Performance varies with the computation load, which is a remarkable feature of GPU different from CPU. Therefore, compared with the CPU, the scalability performance of GPU is slightly lower.

A computation model with four blocks is assigned to be computed with 4 GPUs in the server to investigate how the topology of GPUs in the server influences the communication efficiency of GPUs. The computation model is shown in Fig. 16a. The size of exchange data between blocks in the horizontal direction is 1/8 of the size of exchange data between blocks in the vertical direction. The topology of GPUs in our test server is similar to type (c) in Fig. 6. There are 10 Titan V GPUs in our server, and each PCIe

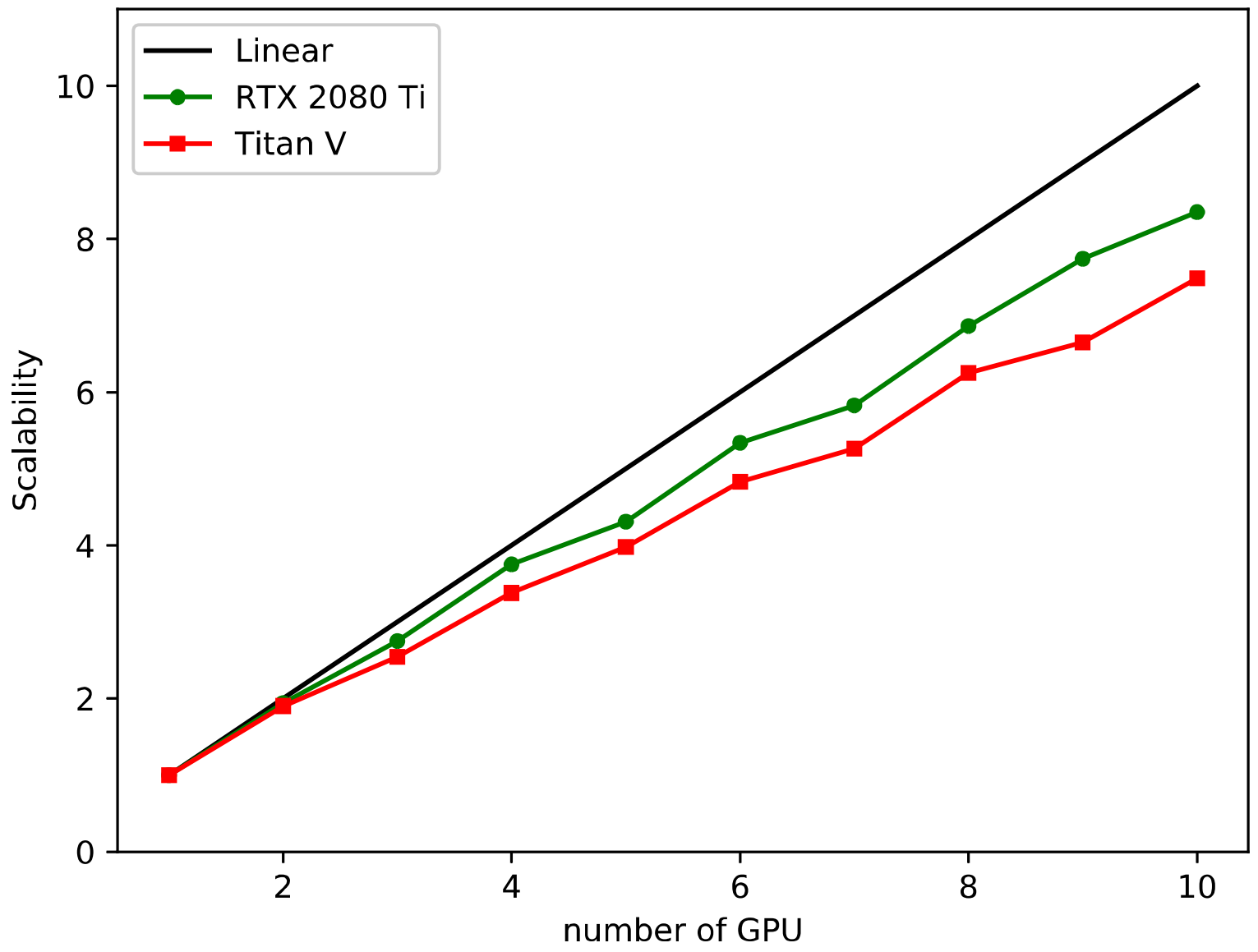

**Fig. 15** Strong scaling result of different GPU. Both GPUs' scalabilities deviate from linear result, but the efficiencies are all larger than 75% when 10 GPUs are utilized. The drops of scalability come from the fact that when grid size decreases, the speedup drops too, see Fig. 13. Scalability of Titan V drops faster than RTX 2080 Ti's, because the speedup of Titan V drops faster than RTX 2080 Ti's when the grid size decreases

switch has 5 GPUs mounted on it. Hence there are two "PIX" GPU groups, GPU 0 $\sim$ 4 and 5 $\sim$ 9. And GPUs between the "PIX" groups are in relation to "PHB". As we have introduced before, GPUs in "PIX" relation communicate faster than GPUs in "PHB" relation. So putting processes that have large data transfer between each other in "PIX" group of GPUs will get higher communication efficiency. For better comparison, five cases are considered, as shown in Fig. 16b. For all cases, the blocks with the same color mean that they are in the same PCIe switch. Case I and case II are used to test the communication performance of GPUs in each PCIe switch. In case III, processes with the largest exchange data are not distributed to GPUs in the same PCIe. In case IV, processes with the largest exchange data are distributed to GPUs in the same PCIe, and it is expected to get higher efficiency than case III. Case V is designed to be the worst situation. We only run the test on a server with Titan V GPU because RTX 2080 Ti GPU does not support P2P communication technique without Nvlink, while Titan V supports P2P by PCIe. The result is shown in Fig. 17. As expected, cases running in the same PCIe switch are the best. Case IV also gets good performance as the cases I and II due to reasonable distribution of communication load. Case V gets the worse performance because each GPU communicates with GPUs that are all located in another PCIe switch, leading to the worst distribution of communication load. The results indicate that the GPU-GPU communication optimization technique is crucial for programs running on the current GPU server in order to fully exploit the performance.

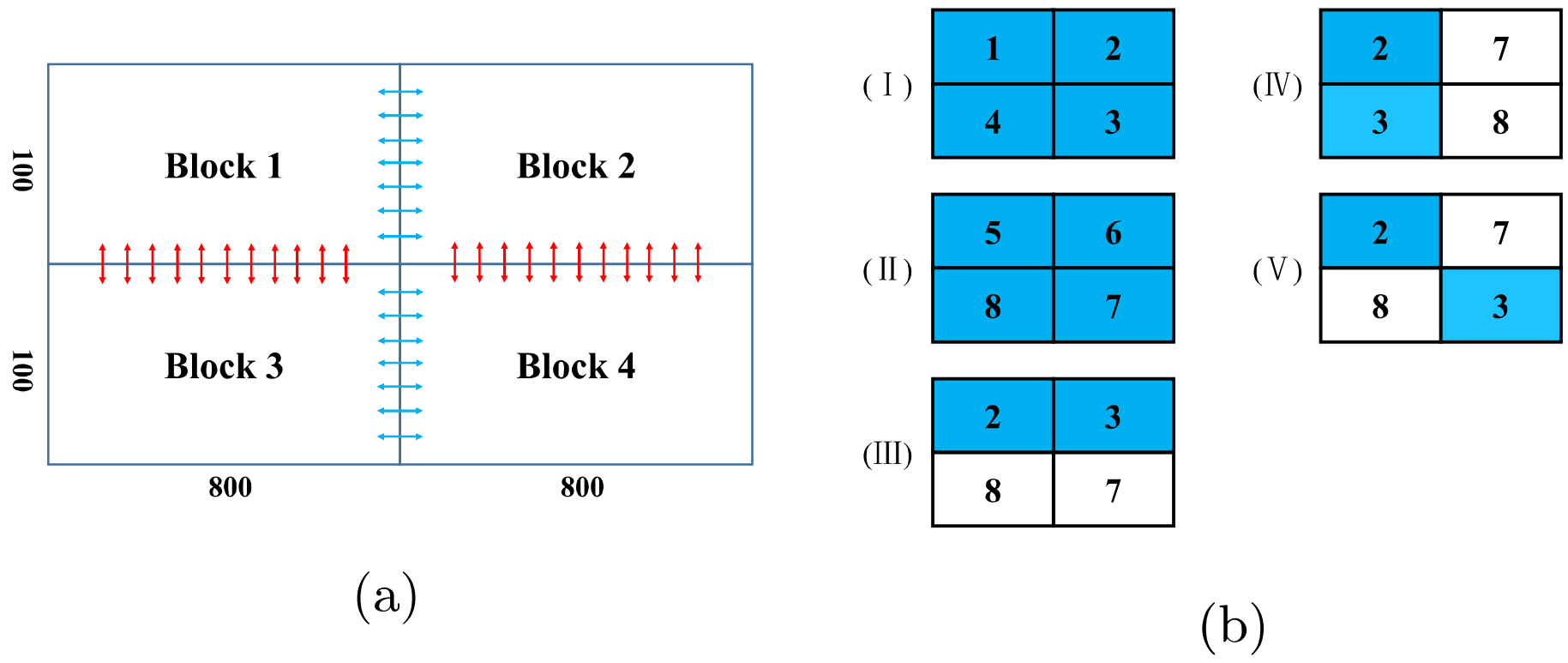

**Fig. 16** Cases configuration. In figure (**a**), there are 100 points in the spanwise direction. Data exchange in vertical direction (red) is larger than data exchange in horizontal direction (blue). In figure (**b**), five cases are set to test the communication performance of different strategies of computations assigned to different GPUs. The number in block presents the GPU ID in which computation is performed. Blocks with the same color indicate that they are in the same PCIe switch. In cases I and II, computations are performed on GPUs that are in two different PCIe switches respectively. In cases III to IV, computations are performed on GPUs belonging to both two PCIe switches. For case III, the maximum data transfers are performed across the PCIe switch. In case IV, the maximum data transfers are performed within the same PCIe switch. In case V, all data transfers are performed across the PCIe switch

## 6 Case study

To further investigate the performance of our solver in the practical application, a planar supersonic jet is simulated with implicit large eddy simulation (ILES). The configuration is similar to the Ref. [16], and we simulated an under-expanded planar jet with the ratio of pressure $p_e/p_\infty = 2.09$. The computational domain is discretized by a Cartesian grid with a total of 67.7 million points. The physical domain, which excludes the sponge zone, has dimensions $64h \times 30h \times 5h$, with a nozzle extending over $0.6h$ inside the domain. The jet height is $h = 3.5mm$ corresponding to a Reynolds number based on the jet height and acoustic speed $Re = \frac{a_c h}{\nu} = 8.15 \times 10^4$. The velocity profile inside the nozzle is a laminar Blasius solution with a boundary layer thickness of $\delta = 0.05h$, and there are 12 points within the boundary layer, 70 points inside the jet height. The constant time step $\triangle t = 2.5 \times 10^{-3}$ is implemented. The non-dimensional time for the whole computation is 360, and only the last 210 is outputted for analysis. Three Titan V GPUs are utilized for computation, and the execution time for the simulation is only about 41 hours.

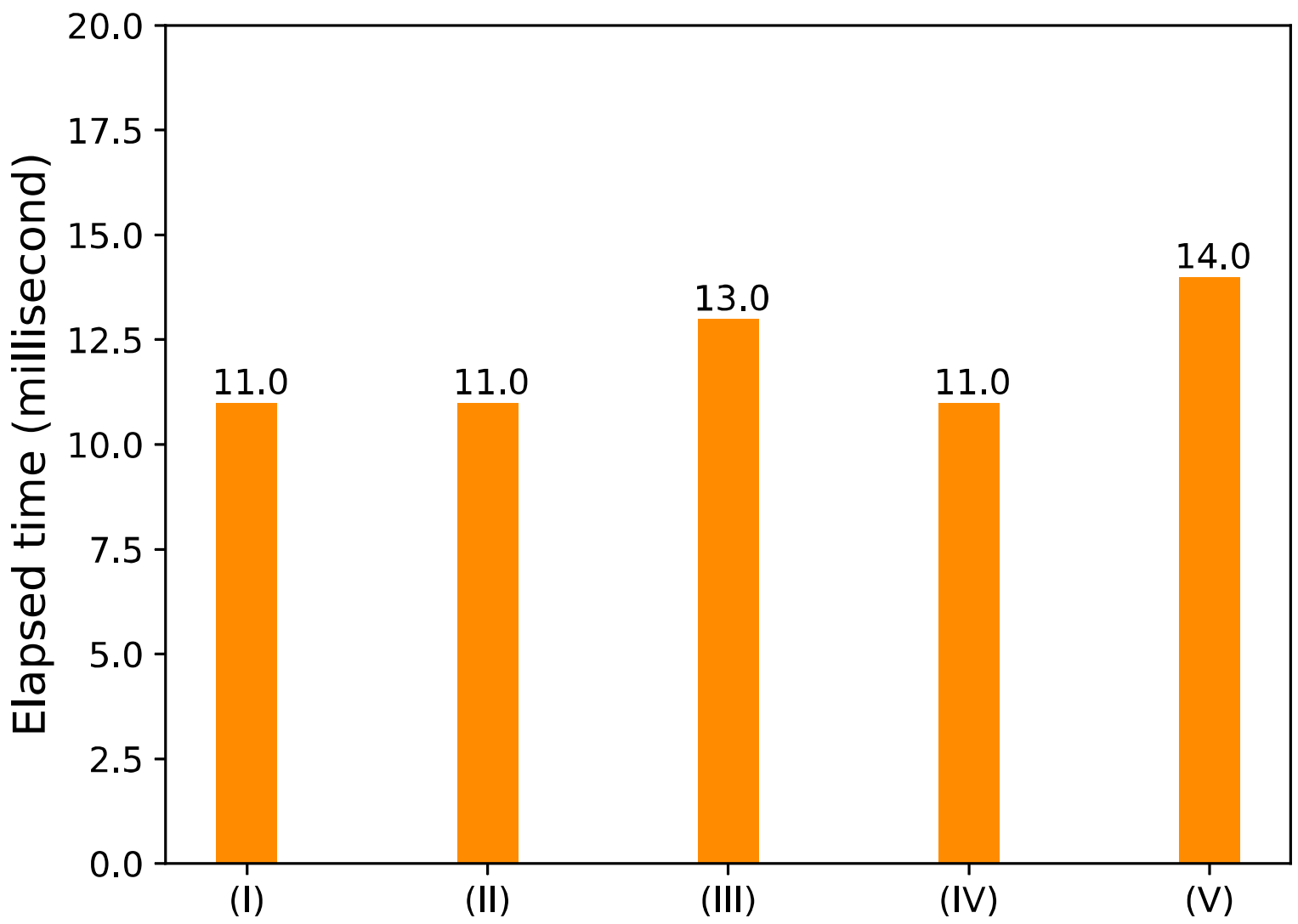

**Fig. 17** Performance of ***HiResX*** running on different GPU topology. Cases in which all GPUs are in the same PCIe switch get best communication efficiency, see case I and II. For the case in which GPUs are located in different PCIe switches, if GPU devices are optimally assigned according to communication load, communication efficiency can also be improved, see case IV. Without GPU to GPU optimization, the communication efficiency is lower, see case III and V

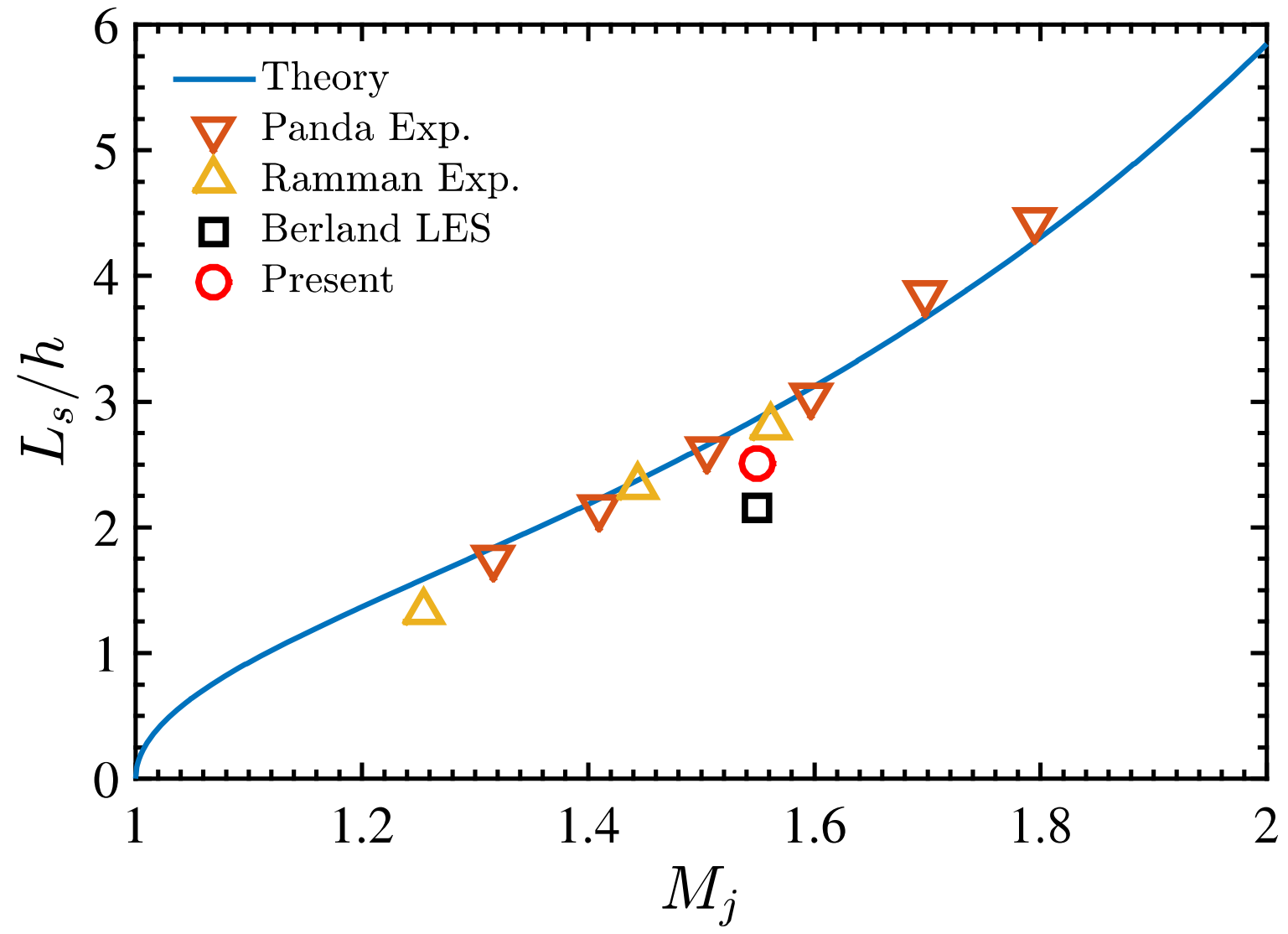

**Fig. 18** Shock-cell spacing $L_h/h$ as a function of the fully expanded jet Mach number $M_j$. Our result is more close to theoretical [17] and experimental results [18, 19], compared to the LES results of Berland et al. [16]

Figure 18 shows the shock-cell spacing $L_s/h$ as a function of the fully expanded jet Mach number $M_j$, and various experimental results [18, 19], LES data of Berland et al. [16] and theoretical data [17] are plotted for comparison for rectangular jets. The present result is closer to experimental results than that of Berland et al.. In Fig. 19, the Strouhal number of the fundamental discrete frequency is given and plotted against the fully expanded jet Mach number for several experiments on rectangular jets, where the experimental data of Panda [18] and LES data of Berland et al. [16] are shown for comparison. Our result is in good agreement with the experimental result. Figure 20 illustrates the flow structure of the jet and its generated acoustic field, which are presented with spanwise vorticity $\omega_z$ and dilatation in the plane $z/h = 2.5$. More details can be found in Ref. [20]. The above results prove the reliability of our solver, and the short execution time demonstrates the high efficiency of our solver running with GPUs.

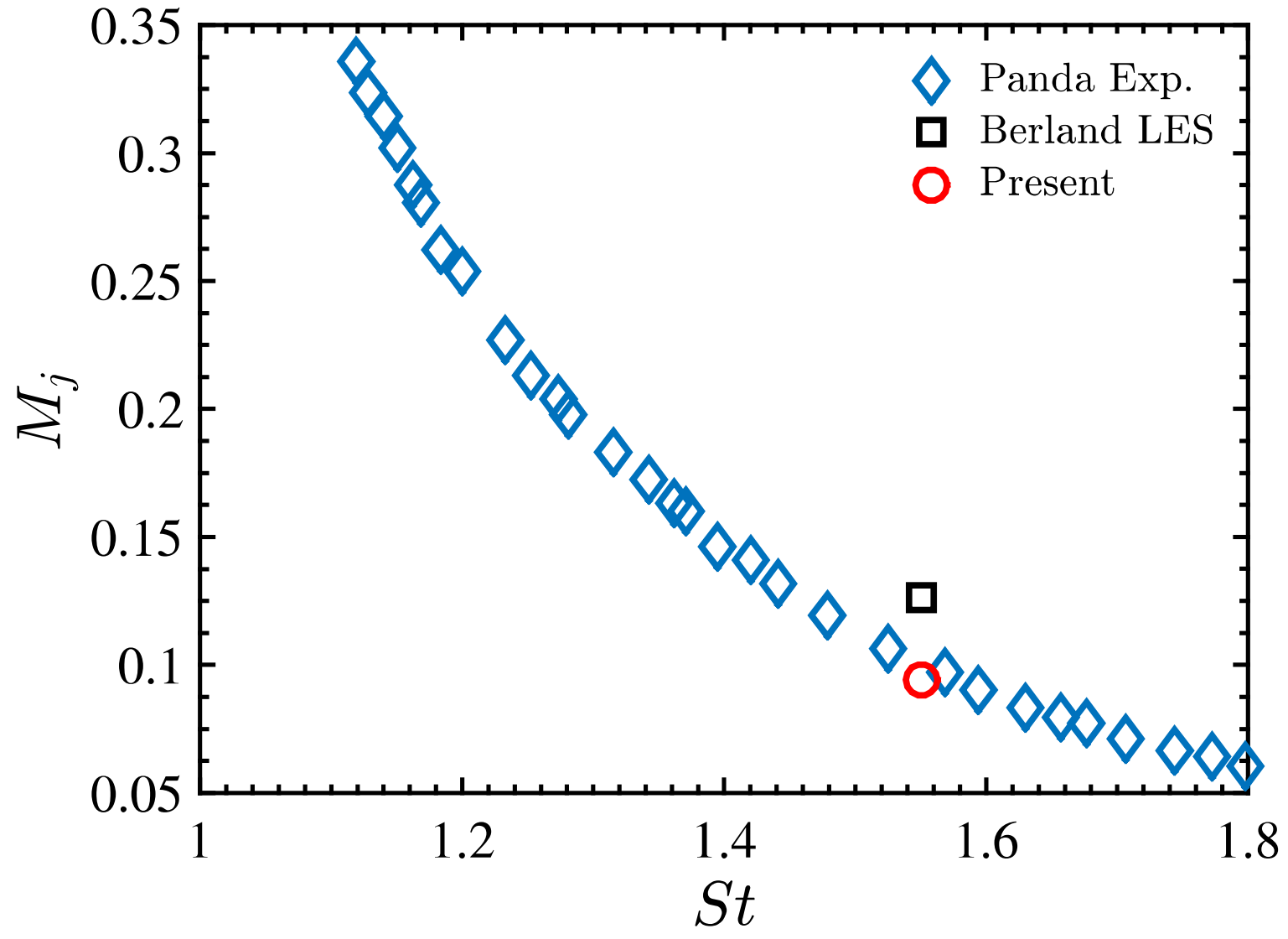

**Fig. 19** Strouhal number $St = fh/Uj$ of the fundamental screech tone as a function of the fully expanded jet Mach number $M_j$. Our result is in good agreement with the experimental result of Ref. [18], which is better than the LES result of Ref. [16]

## 7 Conclusions

Heterogeneous computing is changing our way of scientific computing. The boom of GPU computing in the past several years shows the power and potential of GPU on computing, and it attracts more and more researchers to exploit its application in their fields, including CFD simulation. In the early years of GPU computing, many attempts have been made to port CFD codes onto GPU, but the results are not as satisfying as expected. With the development of GPU computing, more powerful GPUs and related technologies are present, prompting us to further exploit its application in CFD.

In this work, we analyze the characteristics of architectures of current GPU servers and propose a set of techniques to improve the efficiency of data transfer between CPU and GPU and efficiency of communication between GPUs. An in-house compressible flow solver based on high order finite difference method on curvilinear coordinates is successfully ported to GPU with CUDA C. By careful memory planning, we save unnecessary data transfer between CPU and GPU without significant data sacrifice of the capability of grid size of our solver. The “atomic operation” technique and kernel decomposition technology are applied to design high-efficiency kernels. Tests have shown that our solver gets maximum speedups of almost 2000x on a Titan V GPU, 1500x on a Tesla P100, and 650x on an RTX 2080 Ti GPU over a CPU core E5-2680v3. The hardware-aware technology is proved to be more efficient than unoptimizable scenarios. A test case of a supersonic jet

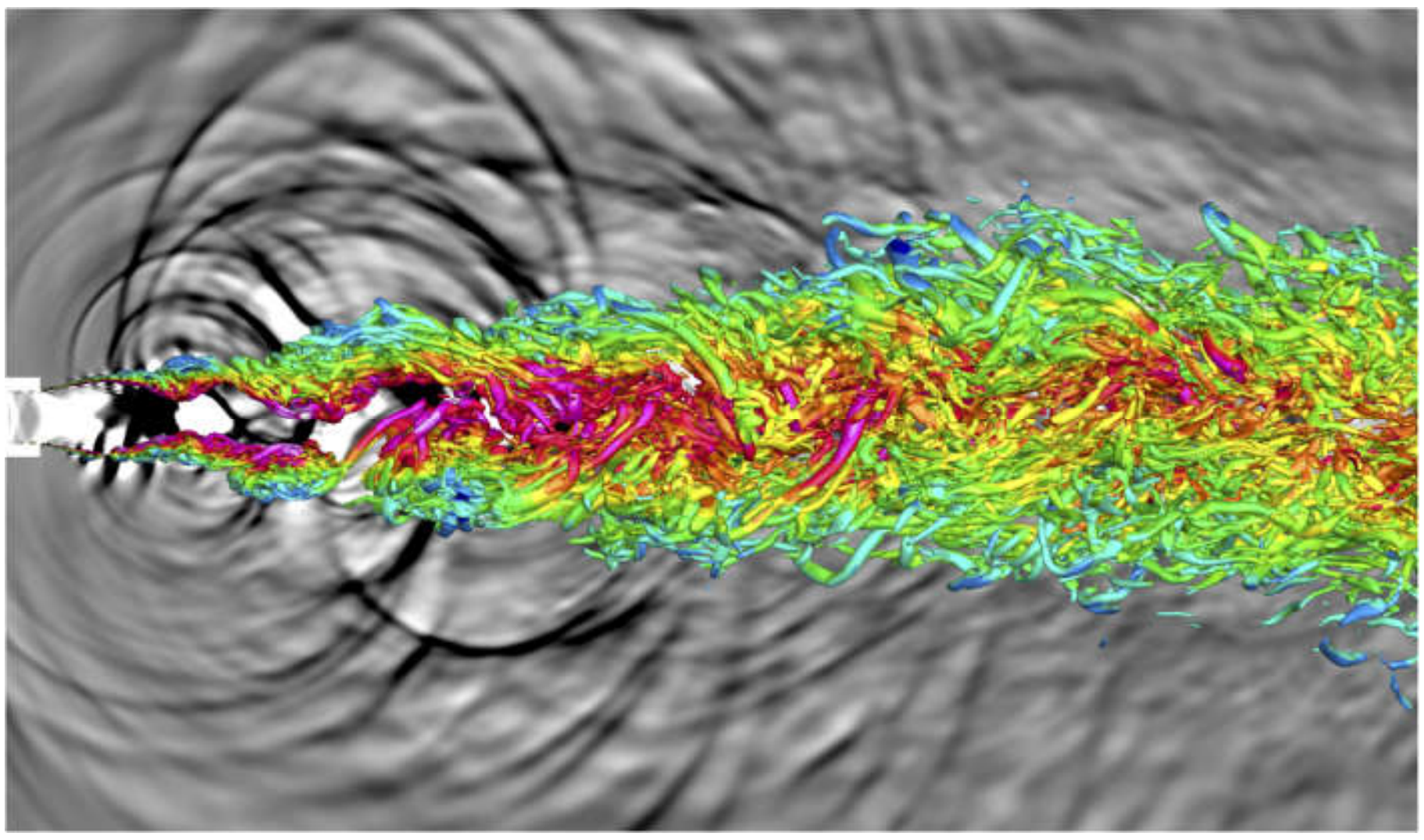

**Fig. 20** Instantaneous snapshot of spanwise vorticity $\omega_z$ and the dilatation in the plane $z/h = 2.5$ as the background. The isosurface of vorticity is colored with the amplitude of velocity

shows the practical application ability of the present solver. This work provides a systematic and efficient solution to apply GPU computing in CFD simulation with specific high order finite difference methods on current GPU heterogeneous computers.

**Acknowledgements**
The authors thank the Supercomputing Center of University of Science and Technology of China for providing powerful computation resources.

**Authors' contributions**
C.Y.: Methodology, Code development, Analysis - Draft writing. P.Z.: Hardware platform construction, Numerical simulation, Data analysis. Z.W.: Supervision, Funding acquisition, Writing - Review & edit. R.Y.: Constructive suggestions - Review. D.S.: Supervision, Funding acquisition - Review. The authors read and approved the final manuscript.

**Funding**
This work is supported by the National Numerical Windtunnel Project, the National Natural Science Foundation projects (91952103,11772323,11621202) and the Fundamental Research Funds for the Central Universities.

**Availability of data and materials**
The data that support the findings of this study are available from the corresponding author upon reasonable request.

**Competing interests**
The authors declare that they have no competing interests.